\documentclass[12pt]{tcibook}
\usepackage{fancyhea}
\usepackage{work}
\usepackage{bm}       
\usepackage{graphicx}
\usepackage{multirow}
\usepackage{lineno}
\usepackage{threeparttable}
\usepackage[normalem]{ulem}
\usepackage{color}
\usepackage[colorlinks = true,
            linkcolor = blue,
            urlcolor  = blue,
            citecolor = blue,
            anchorcolor = blue]{hyperref}

\usepackage{enumitem}
\usepackage[sort&compress,comma,numbers]{natbib}
\usepackage{xspace}
\usepackage{amsmath,amssymb}
\usepackage{comment}

\def\beq{\begin{equation}}
\def\eeq#1{\label{#1}\end{equation}}
\def\eeqn{\end{equation}}

\newenvironment{Eqnarray}%
   {\arraycolsep 0.14em\begin{eqnarray}}{\end{eqnarray}}
\def\beqa{\begin{Eqnarray}}
\def\eeqa#1{\label{#1}\end{Eqnarray}}
\def\eeqan{\end{Eqnarray}}

\let\bar=\overbar

\def\lsim{\mathrel{\raise.3ex\hbox{$<$\kern-.75em\lower1ex\hbox{$\sim$}}}}
\def\gsim{\mathrel{\raise.3ex\hbox{$>$\kern-.75em\lower1ex\hbox{$\sim$}}}}

\def\del{\partial}
\def\Dslash{\not{\hbox{\kern-4pt $D$}}}
\def\dslash{\not{\hbox{\kern-2pt $\del$}}}
\def\pslash{\not{\hbox{\kern-2pt $p$}}}
\def\ETmiss{\not{\hbox{\kern-4pt $E$}}_T}

\def\Dlr{\mathrel{\raise1.5ex\hbox{$\leftrightarrow$\kern-1em\lower1.5ex\hbox{$D$}}}}

\def\MSB{{\bar{M \kern -2pt S}}}
\def\msb{{\bar{\scriptsize M \kern -1pt S}}}

\def\drb{{\bar{\scriptsize D \kern -1pt R}}}

\def\authorlist#1#2{
    \vskip 0.4in
\begin{center}\begin{large} {\bf  #1 } \end{large}
    \vskip 0.2in
              #2
     \vskip 0.2in
   \end{center}
}

\begin{document}


\pagenumbering{roman}

\parindent=0pt
\parskip=8pt
\setlength{\evensidemargin}{0pt}
\setlength{\oddsidemargin}{0pt}
\setlength{\marginparsep}{0.0in}
\setlength{\marginparwidth}{0.0in}
\marginparpush=0pt


\pagenumbering{arabic}

\renewcommand{\chapname}{chap:intro_}
\renewcommand{\chapterdir}{.}
\renewcommand{\arraystretch}{1.25}
\addtolength{\arraycolsep}{-3pt}

\definecolor{offblue}{rgb}{0.26,0.4,0.74}
\definecolor{dkred}{rgb}{0.5,0,0}
\definecolor{dkgreen}{rgb}{0.004, 0.267, 0.129}

\newcommand{\COMMENT}[3]{\textcolor{#1}{(#2: #3)}}
\newcommand{\FIXME}[1]{\textcolor{red}{\bf #1}}

\newcommand{\ADW}[1]{\COMMENT{offblue}{ADW}{#1}} 
\newcommand{\CPW}[1]{\COMMENT{dkred}{CPW}{#1}} 
\newcommand{\HBY}[1]{\COMMENT{dkgreen}{HBY}{#1}} 

\newcommand{\LCDM}{\ensuremath{\Lambda{\rm CDM}}\xspace}
\newcommand{\Lya}{Ly-$\alpha$\xspace}
\newcommand{\Gaia}{\ensuremath{Gaia}\xspace}
\newcommand{\roughly}{\ensuremath{ {\sim}\,} }

\setcounter{chapter}{2} 
\setcounter{tocdepth}{2}


\chapter{Cosmic Probes of Dark Matter}

\authorlist{Conveners: Alex Drlica-Wagner, Chanda Prescod-Weinstein, Hai-Bo Yu}
{Contributors: 
Andrea~Albert,
Mustafa~Amin,
Arka~Banerjee,
Masha~Baryakhtar,
Keith~Bechtol,
Simeon~Bird,
Simon~Birrer,
Torsten~Bringmann,
Regina~Caputo,
Sukanya~Chakrabarti,
Thomas~Y.~Chen,
Djuna~Croon,
Francis-Yan~Cyr-Racine,
William~A.~Dawson,
Cora~Dvorkin,
Vera~Gluscevic,
Daniel~Gilman,
Daniel~Grin,
Ren\'ee~Hlo\v{z}ek,
Rebecca~K.~Leane,
Ting~S.~Li,
Yao-Yuan~Mao,
Joel~Meyers,
Siddharth~Mishra-Sharma,
Julian~B.~Mu\~noz,
Ferah~Munshi,
Ethan~O.~Nadler,
Aditya~Parikh,
Kerstin~Perez,
Annika~H.~G.~Peter,
Stefano~Profumo,
Katelin~Schutz,
Neelima~Sehgal,
Joshua~D.~Simon,
Kuver~Sinha,
Monica~Valluri,
Risa~H.~Wechsler
}


\begin{center}
\vspace{1cm} \noindent{\bf \large Abstract} \\[+1em]
\begin{minipage}{.85\linewidth}
Cosmological and astrophysical observations currently provide the only robust, positive evidence for dark matter. 
Cosmic probes of dark matter, which seek to determine the fundamental properties of dark matter through observations of the cosmos, have emerged as a promising means to reveal the nature of dark matter. 
This report summarizes the current status and future potential of cosmic probes to inform our understanding of the fundamental nature of dark matter in the coming decade.
\end{minipage}
\end{center}

\begingroup
\let\cleardoublepage\clearpage
\tableofcontents
\endgroup

\section{Executive Summary}
\label{sec:cf3summary}

The existence of dark matter, which constitutes $\roughly 85\%$ of the matter density and $\roughly 26\%$ of the total energy density of the universe, is a clear demonstration that we lack a complete understanding of fundamental physics. The  importance, impact, and interdisciplinary nature of the dark matter problem make it one of the most exciting questions in science today. In this report, we explain how cosmic probes of dark matter, which determine microscopic properties of dark matter through cosmic observations on macroscopic scales, have emerged as one of the most promising ways to unveil the nature of dark matter. The HEP community must seize the opportunity to fully realize the potential of cosmic probes of dark matter and broaden its approach to the dark matter problem accordingly in the coming decade.

Cosmological and astrophysical observations currently provide the {\it only} robust, empirical evidence for dark matter. These observations, together with numerical simulations of cosmic structure formation, have established the cold dark matter (CDM) paradigm and lay the groundwork for all other experimental efforts to characterize the fundamental properties of dark matter. In the coming decade, we will measure the distribution of dark matter in the cosmos with unprecedented precision. In particular, we are at the threshold of definitively testing the predictions of CDM on galactic and sub-galactic scales, and any observed deviation would revolutionize our understanding of the fundamental nature of dark matter. Furthermore, by accessing extreme scales and environments, cosmic probes are sensitive to very rare interactions between dark matter and normal matter that are inaccessible in conventional dark matter searches. Cosmic measurements of dark matter properties are entering the {\it precision era}.

This report summarizes the ways in which cosmic probes have emerged as a new field in the endeavor to measure the fundamental, microscopic properties of dark matter. We first identify three core HEP community priorities for cosmic probes of dark matter:

\begin{itemize}[nosep]
    \item Current/near-future HEP cosmology experiments have direct sensitivity to dark matter particle physics \citep[][]{Valluri:2022nrh,Mao:2022fyx,Dvorkin:2022bsc}. {Cosmological studies of dark matter should be supported as a key component of the HEP Cosmic Frontier program due to their unique ability to probe dark matter microphysics and link the results of terrestrial dark matter experiments to cosmological measurements.}

    \item {{The construction of future cosmology experiments is critical for expanding our understanding of dark matter physics.} Proposed facilities across the electromagnetic spectrum, as well as gravitational waves, can provide sensitivity to dark matter physics \citep[][]{Chakrabarti:2022cbu}. The HEP community should make strategic investments in the design, construction, and operation of these facilities in order to maximize their sensitivity to dark matter physics.}

    \item Cosmic probes provide robust sensitivity to the microphysical properties of dark matter due to enormous progress in theoretical modeling, numerical simulations, and astrophysical observations. {Theory, simulation, observation, and experiment must be supported together to maximize the efficacy of cosmic probes of dark matter physics.}
\end{itemize}

We have identified the following  major scientific opportunities for cosmic probes of dark matter physics in the coming decade. These opportunities are summarized below, presented at length in the subsequent sections of this report, and discussed in detail in a set of community white papers~\citep{Bechtol:2022koa, Banerjee:2022qcb,Bird:2022wvk, Berti:2022rwn, Chakrabarti:2022cbu, Valluri:2022nrh, Mao:2022fyx, Dvorkin:2022bsc, Dvorkin:2022pwo, Dienes:2022zbh,Burns:2021pkx}.

\noindent \textbf{Major Scientific Opportunities}

\begin{enumerate}[nosep]

\item \textbf{The HEP community should support measurements of the dark matter distribution as a key element of its Cosmic Frontier program.} The Standard Model of particle physics and cosmology can be tested at unprecedented levels of precision by measuring the cosmic distribution of dark matter. These measurements span an enormous range of scales from the observable universe to sub-stellar-mass systems
(e.g., the matter power spectrum, the mass spectrum of dark matter halos, dark matter halo density profiles, and abundances of compact objects)
\citep{Bechtol:2022koa,Bird:2022wvk,Brito:2022lmd}. Novel particle properties of dark matter (e.g., self-interactions, quantum wave features, tight couplings with radiation) can lead to observable signatures in the distribution of dark matter that differ from the CDM prediction.

\item \textbf{The HEP community should pursue the detection of dark matter halos below the threshold of galaxy formation as an exceptional test of fundamental dark matter properties.} The CDM model makes the strong, testable prediction that the mass spectrum of dark matter halos extends below the threshold at which galaxies form, $\mathcal{O}(10^7)~M_{\odot}$~\citep{Bechtol:2022koa}. 
Sub-galactic dark matter halos are less influenced by baryonic processes making them especially clean probes of fundamental physics of dark matter. 
We are on the cusp of detecting dark matter halos that are devoid of visible stars through several cosmic probes (e.g., strong lensing and the dynamics of stars around the Milky Way). 

\item \textbf{Instruments, observations, and theorists that study extreme astrophysical environments should be supported as an essential way to explore the expanding landscape of dark matter models.} Extreme astrophysical environments provide unique opportunities to explore dark matter couplings to the Standard Model that are inaccessible with conventional dark matter search experiments and span ${\sim}50$ orders of magnitude in dark matter particle mass~\citep{Berti:2022rwn}.

\item \textbf{HEP computational expertise and resources should be combined with astrophysical simulation expertise to rapidly advance numerical simulations of dark matter physics.} Numerical simulations of structure formation and baryonic physics are critical to robustly interpret observational results and disentangle new dark matter physics from astrophysical effects~\citep{Banerjee:2022qcb}. Simulations are thus essential to address particle physics questions about the nature of dark matter. 

\item \textbf{While large experimental collaborations are naturally matched to the HEP community, new mechanisms are also needed to support emerging, interdisciplinary efforts.} The interdisciplinary nature of cosmic probes of dark matter calls for innovative new ways to support the pursuit of scientific opportunities across traditional disciplinary boundaries. Sustained collaboration between particle theorists, gravitational dynamicists, numerical simulators, observers, and experimentalists is required to fully realize the power of cosmic probes of dark matter.

\end{enumerate}

These major opportunities represent pathways for transforming our understanding of dark matter. We observe that the trend in dark matter research has changed profoundly in the last 10 years. In particular, particle physicists and astrophysicists have collaborated to make a series of breakthroughs in studying dark matter theories beyond the CDM paradigm. For example, the idea of dark sectors, i.e., that dark matter may reside in its own sector and carry its own forces, has been firmly established, together with associated theoretical tools. Such a dark force could operate at the most fundamental level with a range of $\mathcal{O}(10^{-12})~{\rm cm}$, but it would change the dark matter distribution within $\mathcal{O}(10^{22})~{\rm cm}$ in galactic halos. Simulators can now perform cosmological hydrodynamical simulations of structure formation that account for the particle physics properties of dark matter, such as self-interactions~\cite{Vogelsberger:2014pda,Robles:2017,Robertson:2017mgj}, quantum wave features~\cite{Mocz:2019pyf}, and tight couplings with radiation~\cite{Lovell:2017eec}. These studies set the basis for quantifying astrophysical uncertainties and disentangling dark matter physics from baryon physics, a key step for extracting the Lagrangian parameters that describe a particular dark matter model. 
Advances since the last Snowmass study combined with the scientific outlook assembled in this Snowmass study motivate the possibility that cosmic probes may result in a transformational change in our understanding of dark matter in the coming decade.

On the experimental side, new observational facilities that directly measure dark matter in the cosmos will be taking data within the next decade, see~\cite{Chakrabarti:2022cbu} for details. For example, the DOE- and NSF-supported Rubin Observatory Legacy Survey of Space and Time (LSST) is scheduled to start full operations in 2024 \citep{LSST:2008ijt}. The primary HEP interest in Rubin LSST is the study of dark energy; however, it has enormous potential to discover new physics beyond CDM \citep{Drlica-Wagner:2019mwo, Mao:2022fyx}. Rubin LSST will observe the faintest satellite galaxies of the Milky Way, stellar streams, and strong lens systems to detect the smallest dark matter halos, thereby probing the minimum mass of ultra-light dark matter and thermal warm dark matter. Gravitational lensing will provide precise measurements of the densities and shapes of dark matter halos, which are sensitive probes of interactions in the dark sector. Microlensing measurements will directly probe primordial black holes and the compact object fraction of dark matter at the sub-percent level over a wide range of masses. On a longer time horizon, the CMB Stage-4 (CMB-S4) project \citep{CMB-S4:2016ple,CMB-S4:2022ght}, another cosmology experiment supported by DOE and NSF, has access to rich dark matter physics as well \citep{Dvorkin:2022bsc}. For instance, detection of additional relativistic degrees of freedom in the early universe would immediately imply the existence of a dark sector. The HEP community strongly supports the inclusion of dark matter physics in the research programs of these experiments alongside studies of dark energy and inflation.

Since cosmic probes of dark matter are multidisciplinary {\it and} interdisciplinary, we must develop new ways to cultivate and sustain collaboration, as well as new approaches to mentoring the next-generation of scientists. A program like the DOE Dark Matter New Initiatives is well-suited to support a small-scale collaborative effort from particle physicists and astrophysicists with a well-defined scientific goal. Within upcoming HEP cosmology experiments such as Rubin LSST and CMB-S4, new effort is needed to assemble collaborative teams to study cosmic probes of dark matter physics. Additional support must be provided to enable dark matter as a parallel branch of fundamental physics being pursued by these experiments. Such a program will fully leverage the unprecedented capabilities of these facilities~\cite{Mao:2022fyx}. On large scales, the construction of future cosmology experiments is critical for expanding our understanding of dark matter physics. HEP involvement will be essential for the design and construction of these facilities~\cite{Chakrabarti:2022cbu}, and dark matter physics should be a core component of their scientific missions.

Identifying the nature of dark matter is one of the most important tasks in science. Cosmic probes are the most ``expansive'' (and may be the only viable) approach to understanding fundamental properties of dark matter, because they are valid even if the coupling between dark matter and normal matter is extremely weak, even as weak as gravity. With new observational facilities coming online, plans for future facilities emerging, and tremendous progress being made in numerical simulations and semi-analytical modeling, we are entering a transformative decade in the effort to measure the microscopic properties of dark matter from cosmic observations. In this report, we present the possibilities created by a community commitment to a decade of dark matter.

\section{Introduction}

\begin{figure*}[t]
    \centering
    \includegraphics[width=\textwidth]{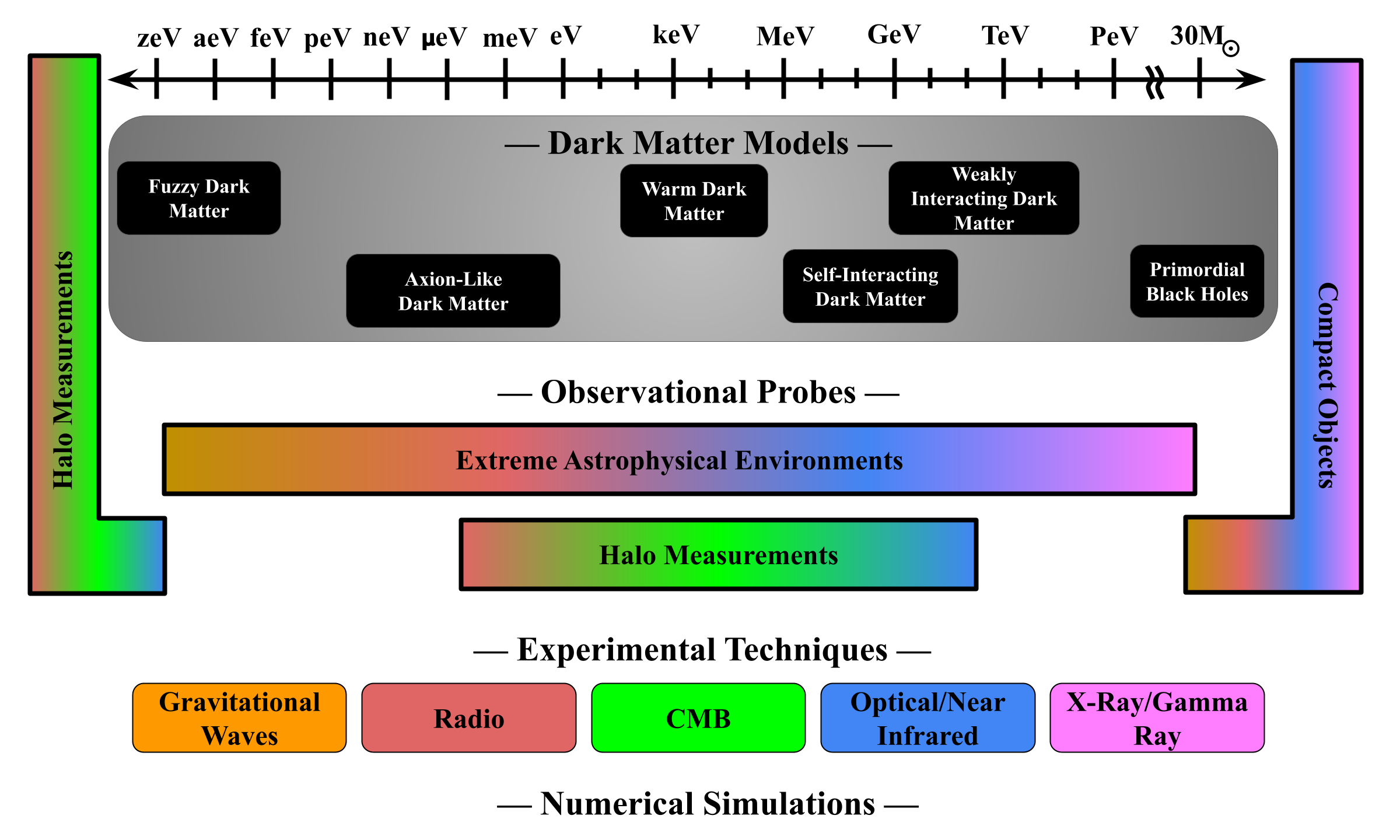}
    \caption{Cosmic observations bound the available dark matter parameter space and probe dark matter physics over the entire allowed mass range. Cosmic probes explore the fundamental physics of dark matter both through gravity alone and through dark matter interactions with the Standard Model. Cosmic probes of  dark matter physics are highly complementary to cosmological measurements of dark energy, inflation, and neutrinos. Furthermore, cosmic probes provide essential information for designing and interpreting terrestrial searches for dark matter. Figure inspired by similar figures in the literature \citep[e.g.,][]{Battaglieri:2017aum, Abazajian:2019eic,Antypas:2022asj}.
    }
    \label{fig:overview}
\end{figure*}

Over the past several decades, experimental searches for non-baryonic dark matter have proceeded along several complementary avenues.
Collider experiments attempt to produce and detect the presence of dark matter particles, while direct detection experiments attempt to measure energy deposition from very rare interactions between dark matter and Standard Model particles.
In parallel, indirect dark matter searches seek to detect the energetic Standard Model products from the annihilation or decay of dark matter particles {\it in situ} in astrophysical systems. 
Despite these extensive efforts, the only positive measurements of dark matter to date come from astrophysical and cosmological observations. 
This report summarizes the exciting scientific opportunities presented by cosmic probes of fundamental dark matter physics in the coming decade, as highlighted in Fig.~\ref{fig:overview} schematically.
The content of this report has been primarily guided by five solicited white papers \citep{Bechtol:2022koa, Banerjee:2022qcb, Bird:2022wvk, Berti:2022rwn, Chakrabarti:2022cbu} and six contributed white papers from the HEP community \citep{Valluri:2022nrh, Mao:2022fyx, Dvorkin:2022bsc, Dvorkin:2022pwo, Dienes:2022zbh, Burns:2021pkx}.

Astrophysical and cosmological observations are a unique, powerful, and complementary technique to study the fundamental properties of dark matter. 
They probe dark matter directly through gravity, the only force to which dark matter is known to couple. 
On large cosmological scales, current observational data can be described by a simple cosmological model containing stable, non-relativistic, collisionless, cold dark matter.
However, many viable theoretical models of dark matter predict observable deviations from CDM that  are testable with current and future experimental programs.
Fundamental physical properties of dark matter---e.g., particle mass, time evolution, self-interaction cross section,  and coupling to the Standard Model or other dark sector particles---can imprint themselves on the macroscopic distribution of dark matter in a detectable manner.

In addition, astrophysical observations complement terrestrial dark matter searches by providing input to direct and indirect dark matter experiments, and by enabling alternative tests of any non-gravitational coupling(s) between dark matter and the Standard Model.  
For example, astrophysical observations are required to (\textit{i}) measure the local density and velocity distribution of dark matter, an important input for designing and interpreting direct dark matter searches, 
(\textit{ii}) identify and characterize regions of high dark matter density, an important input for targeting and interpreting indirect searches, and 
(\textit{iii}) set strong constraints on the particle properties  of dark matter, an important input for designing novel terrestrial dark matter experiments with viable discovery potential.  
In the event of a terrestrial dark matter detection---e.g., the detection of a weakly interacting massive particle (WIMP) or axion---cosmic observations will be crucial to interpret terrestrial measurements in the context of cosmic dark matter.
Furthermore, cosmic probes provide critical information to direct future terrestrial searches for novel dark matter candidates.  
Finally, in many cases, astrophysical and cosmological observations provide the \emph{only} robust constraints on the viable range of dark matter models.

There is also immense discovery potential at the intersection of particle physics, cosmology, and astrophysics.
The detection of dark matter halos that are completely devoid of visible galaxies would provide an extremely sensitive probe of new dark matter physics.
Measuring a deviation from the gravitational predictions of CDM in these halos would provide much-needed experimental guidance on dark matter properties that are not easily measured in particle physics experiments (e.g., dark matter self-interaction cross sections). 
Likewise, results from terrestrial particle physics experiments can suggest specific deviations from the CDM paradigm that can be tested with astrophysical observations.
The expanding landscape of theoretical models for dark matter strongly motivates the exploration of dark matter parameter space beyond the current sensitivity of the HEP program.

In fact, cosmology has a long history of testing the fundamental properties of dark matter. 
For instance, neutrinos were long considered a viable candidate to make up all the dark matter \citep[e.g.,][]{Schramm:1981,Kolb:1988}.
The 30\,eV neutrino dark matter candidate of the 1980s is an especially interesting case study of the interplay between particle physics experiments and astrophysical observations.
In 1980, Lubimov et al.\ reported the discovery of a non-zero neutrino rest mass in the range $14\,{\rm eV} < m_{\nu} < 46\,{\rm eV}$ \citep{Lubimov:1980}. 
Neutrinos with this mass would provide a significant fraction of the critical energy density of the universe, but would be relativistic at the time of decoupling, thus manifesting as hot dark matter. 
Over the next decade, this ``discovery'' was aggressively tested by several other tritium $\beta$-decay experiments.
During this same period, the first measurements of the stellar velocity dispersion of dwarf spheroidal galaxies showed that these galaxies are highly dark matter dominated. 
The inferred dark matter density within the central regions of dwarf galaxies was used to place lower limits on the neutrino rest mass that were incompatible with the 30\,eV neutrino dark matter candidate \citep{Aaronson:1983,Gerhard:1992}. 
Furthermore, numerical simulations of structure formation demonstrated that large-scale structure observations are incompatible with a universe dominated by hot dark matter in the form of neutrinos~\citep{White:1983}.
Similar stories can be told of stable heavy leptons and other dark matter candidates that have been excluded by cosmological and astrophysical measurements \citep{Benvenuti:1977a,Lee:1977,Gunn:1978}.
Cosmology has continually shown that the microscopic physics governing the fundamental nature of dark matter and the macroscopic distribution of dark matter are intimately intertwined.

The strong connection between cosmology, astrophysics, and particle physics serve as the motivation for the Dark Matter: Cosmic Probes Topical Group (CF3) within the Snowmass Cosmic Frontier. 
CF3 focuses on the use of cosmological techniques and astrophysical observations to study the fundamental nature of dark matter over the full range of allowed dark matter masses.
While many experimental studies of dark matter search for a previously undetected interactions between dark matter and Standard Model particles, CF3 also seeks to measure the behavior of dark matter ever more precisely in order to compare against the predictions of \LCDM. 
Thus, some of the scientific approaches and experimental facilities proposed by CF3 overlap significantly with cosmological studies of dark energy and the early universe.
CF3 discussions took place between November 2020 and July 2022 through a series of meetings that occurred on a roughly bi-weekly cadence.
CF3 received 75 letters of intent,\footnote{\url{https://snowmass21.org/cosmic/dm_probes}} which resulted in the coordinated submission of five solicited white papers: 

\begin{itemize}[nosep]
    \item {\bf WP1} - Dark matter physics from dark matter halo measurements \citep{Bechtol:2022koa}.
    \item {\bf WP2} - Cosmological simulations for dark matter physics \citep{Banerjee:2022qcb}.
    \item {\bf WP3} - Primordial black hole dark matter \citep{Bird:2022wvk}.
    \item {\bf WP4} - Dark matter in extreme astrophysical environments \citep{Berti:2022rwn}.
    \item {\bf WP5} - Observational facilities to study dark matter physics \citep{Chakrabarti:2022cbu}.
\end{itemize}

\noindent These solicited white papers were complemented by six white papers that were contributed directly to CF3 \citep{Valluri:2022nrh, Mao:2022fyx, Dvorkin:2022bsc, Dvorkin:2022pwo, Dienes:2022zbh,Burns:2021pkx} and numerous related white papers that were contributed to other topical groups by the HEP community \citep[e.g.,][]{Brito:2022lmd,Alvarez:2022kut,Abazajian:2022ofy,Dvorkin:2022jyg,Antypas:2022asj}.
Furthermore, we present three cases that highlight how cosmic probes can play a central role in identifying fundamental properties of dark matter and/or providing complementary information for designing search strategies in terrestrial experiments. This report summarizes nearly two years of community input, and its structure largely follows the CF3 solicited community white papers \citep{Bechtol:2022koa,Banerjee:2022qcb,Bird:2022wvk,Berti:2022rwn,Chakrabarti:2022cbu}.

\section{Dark Matter Halo Measurements}


\begin{figure}[t!]
    \centering
    \includegraphics[width=\textwidth]{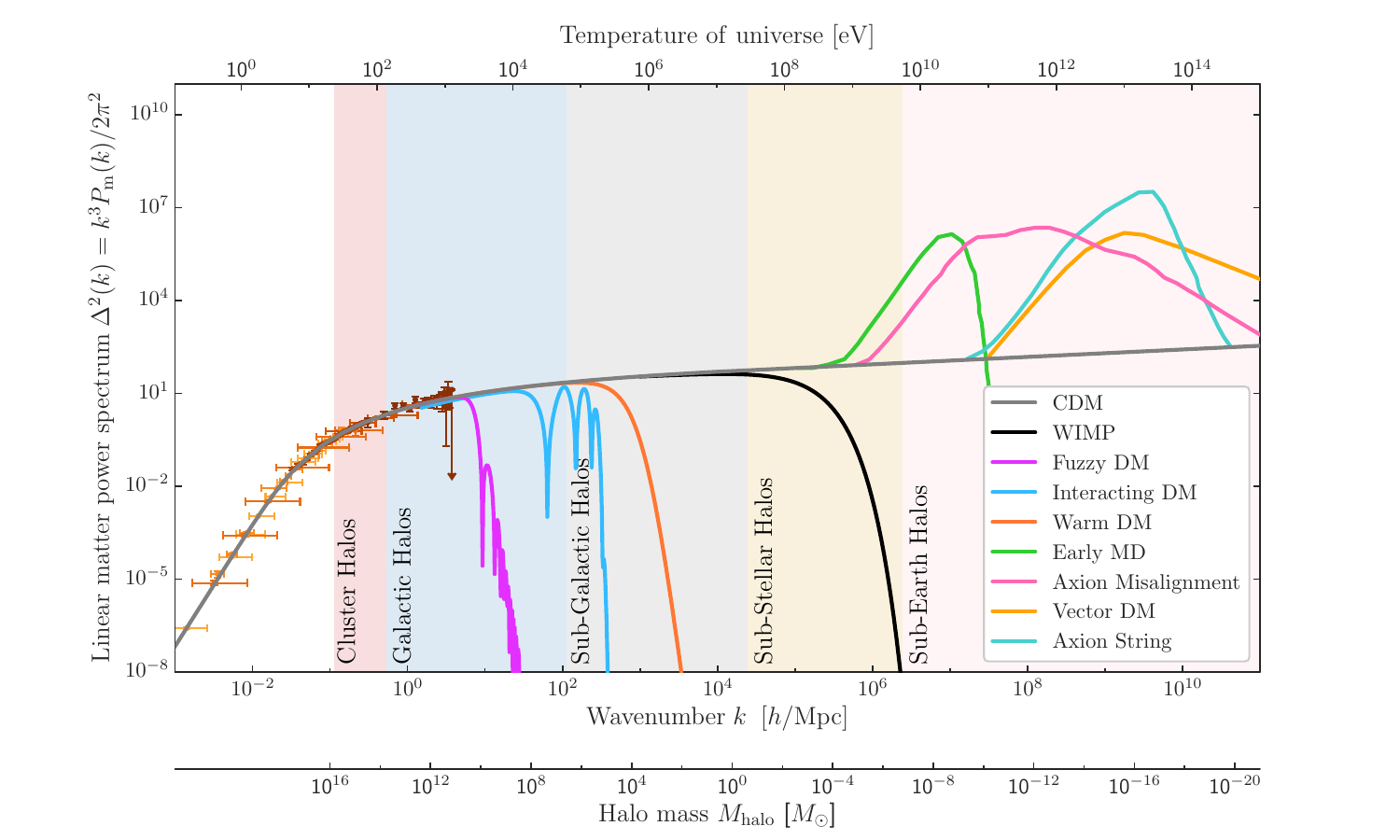}
    \caption{\label{fig:cf3_pk} The dimensionless linear matter power spectrum extrapolated linearly to $z=0$. Theoretical predictions are plotted for four models that suppress power: (1) ultra-light axion ``fuzzy'' dark matter with a mass $m=10^{-22}$~eV (magenta; \cite{Hu:2000ke}), (2) dark matter--baryon interactions with interaction cross section that scales with velocity as $\sigma_0 v^{-4}$ for $\sigma_0 = 10^{-22}$cm$^2$ (blue; \cite{Boddy:2018kfv}), (3) thermal relic warm dark matter with a mass $m\sim 40$~keV (orange; \cite{Viel:2005qj}), (4) weakly interacting massive particle dark matter represented by a bino-like neutralino with a mass, $m\sim 100$~GeV (black; \cite{Green:2005fa}). Also shown are four models that affect power on very small scales: (1) early matter domination assuming a reheat temperature of $10$~MeV (green; \cite{Erickcek:2011us}), (2) post-inflationary production of QCD axions dominated by the misalignment mechanism (pink; \cite{Buschmann:2019icd}), (3) vector dark matter produced during inflation assuming an inflationary scale of $10^{14}$~GeV and a DM mass of $10^{-6}$~eV (gold; \cite{Graham:2015rva}), and (4) post-inflationary production of axions dominated by strings (cyan; \cite{Gorghetto:2020qws}). Note that the position of the power spectrum cutoff and/or enhancement depends on model parameters and is flexible for most cases shown here. Power spectrum measurements on large scales are compiled from \cite{Chabanier:2019eai}. Shaded vertical bands roughly indicate the characteristic kinds of halos formed on each scale, and the horizontal axes indicate wavenumber, halo mass, and the temperature of the Universe when that mode entered the horizon. Figure from~\cite{Bechtol:2022koa}.}
\end{figure}

In the standard model of cosmic structure formation, dark matter in the late-time universe is clustered into gravitationally bound over-densities called halos. These halos provide sites for baryons to cool, collapse, and form galaxies. Astronomical observations show that dark matter halos are distributed according to a power-law mass spectrum extending from the scales of galaxy clusters ($\roughly 10^{15} M_\odot$) to those of ultra-faint dwarf galaxies ($\roughly 10^8 M_\odot$). In the prevailing CDM theory, dark matter is made of collisionless, cold particles or compact objects. The CDM theory does a good job of explaining the large-scale structure of the universe~\cite{Planck:2013pxb} and overall properties of galaxies~\cite{Vogelsberger:2014kha,Hopkins:2017ycn}. However, there are many reasons to believe that CDM is an approximation and that the dark sector is more complex and vibrant.  
From the theory perspective, CDM provides a parametric description of cosmic structure, but it is far from a complete theory. In CDM, the particle properties of dark matter, such as the mass, spin, interaction(s), and production mechanism(s), remain unspecified. 
In fact, many theoretical models describing the particle physics of dark matter predict that the simplest CDM model breaks down at small physical scales~\cite{Bechtol:2022koa}. 
On the observational side, CDM has faced long-standing challenges in explaining detailed measurements of dark matter distributions on galactic and sub-galactic scales~\cite{Tulin:2017ara,Bullock:2017xww}, where we are pushing the boundaries of both observations and numerical simulations.
In the next decade, observations of dark matter halos over a wide range of mass scales will provide unique opportunities to test the vast landscape of dark matter theories and potentially discover deviations from the predictions of CDM.

\begin{figure}[t!]
    \centering
    \includegraphics[width=0.8\textwidth]{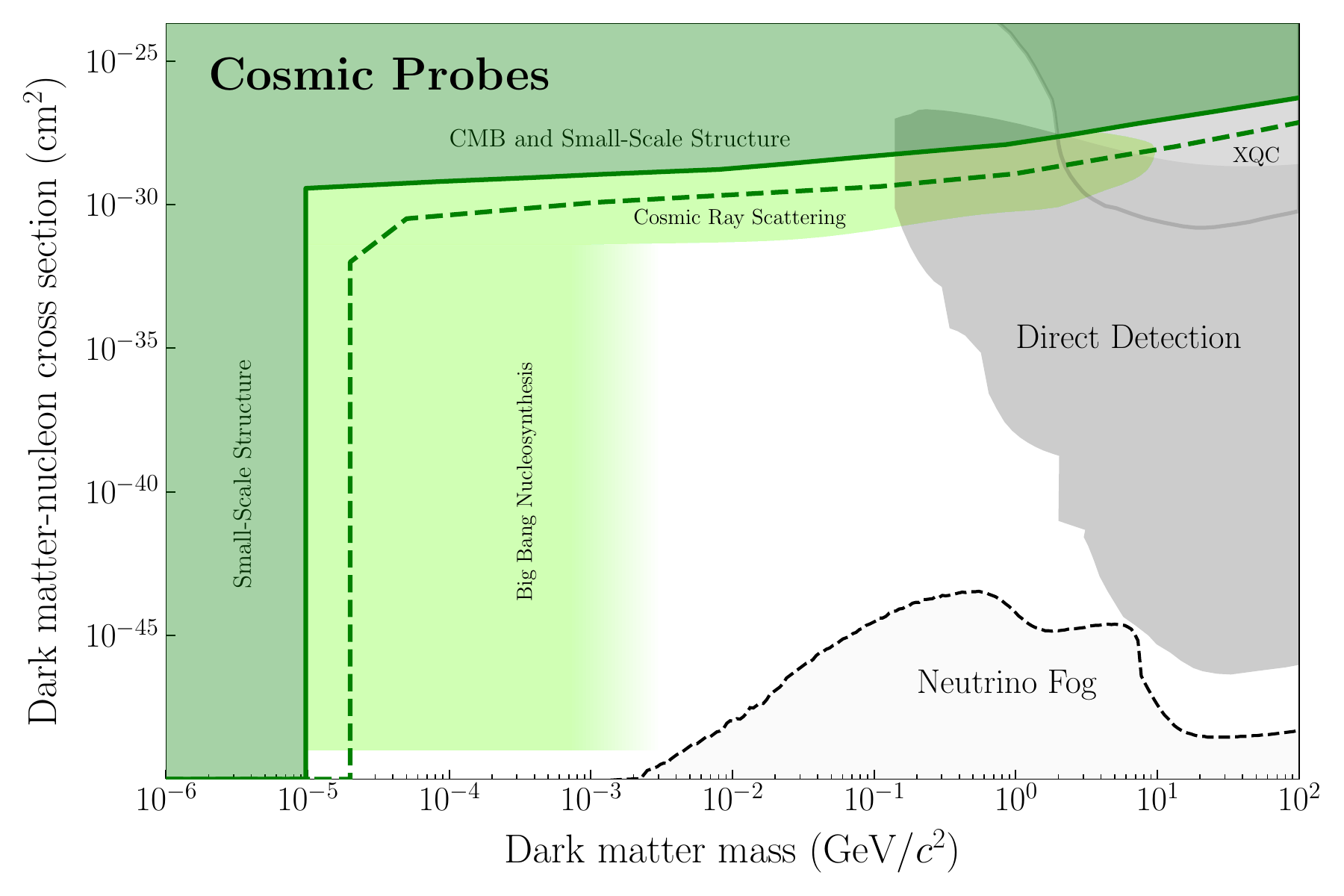}
    \caption{ \label{fig:cf3_idm} Cosmic probes of the matter power spectrum and dark matter halos set strong constraints on the minimum thermal dark matter particle mass \citep[e.g.,][]{Gilman:2019nap, Nadler:2021dft} and spin-independent dark matter--nucleon scattering cross section \citep[e.g.,][]{McDermott:2010pa,Gluscevic:2017ywp,Boddy:2018wzy, Nadler:2019zrb, Rogers:2021byl,Buen-Abad:2021mvc,An:2022sva} (green regions). 
    Projected improvements in sensitivity coming from future facilities and observations are indicated with a dashed green line, based on the estimate with a potential discovery of $10^5~M_\odot$ subhalos using Rubin LSST stream observations.
    Constraints from Big Bang nucleosynthesis \citep[e.g.,][]{Krnjaic:2019dzc} and and cosmic rays upscattering dark matter in the XENON1T experiment \citep[e.g.,][]{Alvey:2022pad} are indicated in light green and are subject to additional model dependence. 
    These constraints are highly complementary to constraints from direct detection experiments \citep[as collected by][]{Akerib:2022ort,Emken:2018run} (gray regions). The neutrino fog for xenon direct detection experiments is shown with dashed black line \citep{OHare:2021utq}. }
\end{figure}

Using halo measurements to study dark matter physics has several advantages.  First, there is a strong connection between dark matter halos and the physics of the early universe. The seeds of cosmological structure formation were established in the earliest moments after the Big Bang. As we measure the distribution of dark matter across a broad range of physical scales, we simultaneously learn about the initial conditions of the universe and probe periods of cosmic history that might be inaccessible by other means. Second, halo measurements are sensitive to a broad range of dark matter models. To date, all positive experimental evidence for the existence and properties of dark matter comes from astrophysical observations. Measurements of the abundance, density profiles, and spatial distribution of dark matter halos offer sensitivity to an enormous range of dark matter models, and are complementary to both terrestrial experiments and indirect searches for dark matter annihilation and decay products. Third, our understanding of how the fundamental properties of dark matter at a microscopic scale impact structure formation throughout cosmic history is rapidly advancing. Recently, there has been tremendous progress in modeling the formation and evolution of dark matter halos in the context of novel dark matter theories beyond CDM. There is enormous potential to further develop detailed phenomenology for a broader range of dark matter models, and to explore new regions of theory space with new and archival data. Thus, halo measurements provide a window into both dark matter physics and early universe cosmology. Fig.~\ref{fig:cf3_pk} illustrates these connections by showing the linear matter power spectrum predicted by several theoretical models of dark matter, together with the relevant scales of the halo mass and temperature of the universe when that mode entered the horizon.
In Fig.~\ref{fig:cf3_idm}, we show the complementarity between measurements of the matter power spectrum and dark matter halos and terrestrial direct detection searches in the context of the spin-independent dark matter--nucleon scattering cross section.

To further leverage halo measurements and extract fundamental properties of dark matter, we set the following observational milestones. First, precision measurements of galaxy-scale dark matter halos are critical. Current and near-future facilities will provide a detailed mapping between luminous galaxies and their invisible dark matter halos across $13$ billion years of cosmic history and more than $8$ orders of magnitude in dark matter halo mass ($10^{15}\,M_{\odot} \gtrsim M_{\rm halo} \gtrsim 10^{7} \,M_{\odot}$). Detailed measurements of halo abundances and density profiles across cosmic time will provide increasingly stringent tests of the CDM paradigm. Second, within the next decade, several observational techniques will become sensitive to dark matter halos at or below the minimum mass required to host stellar populations ($M_{\rm halo} \lesssim 10^{7} \,M_{\odot}$). Population studies of completely dark halos offer unique advantages to study the microphysical properties of dark matter because the evolution of these halos is less affected by baryonic physics. Many theoretical models of dark matter predict conspicuous deviations from CDM in low-mass halos. Third, a suite of innovative and ambitious observational techniques can be used to search for compact stellar- and planetary-mass-scale halos via their subtle gravitational effects ($M_{\rm halo} \lesssim 10^{2}\,M_{\odot}$). 
The discovery of such low-mass halos would immediately transform our understanding of both dark matter properties and the physics of the early universe. 

These observational breakthroughs have the potential to revolutionize our understanding the nature of dark matter. For example, if a cored density profile is inferred in ultra-faint dwarf galaxies, where baryonic feedback is minimal, it will indicate that dark matter may have strong self-interactions or quantum wave features. In this case, combining the measurements of dark matter density profiles from ultra-faint dwarfs to clusters of galaxies, we may narrow down the mass of dark matter particle(s), as well as that of the dark mediator(s). Observations of dark matter halos below the galaxy formation threshold will put strong constraints on the ``warmth'' of dark matter and set upper limits on the interaction strength between dark matter and any warm species, including baryons or dark radiation, in the early universe. On the other hand, if such halos are not detected, it would imply a cutoff in the matter power spectrum and a radical deviation from CDM. Since a cutoff in the matter power spectrum could also reduce the abundance of Milky Way satellite galaxies and affect the Lyman-alpha forest, we could further confirm a departure from CDM with these complementary measurements. Numerical simulations are essential to understand baryonic systematics and to connect the Lagrangian parameters that describe a particle physics model for dark matter to halo observables. Simulations are the topic of the following section.

\section{Cosmological Simulations of Dark Matter}
\label{sec:simulations}

\begin{figure}[t]
    \centering
    \includegraphics[width=0.32\textwidth]{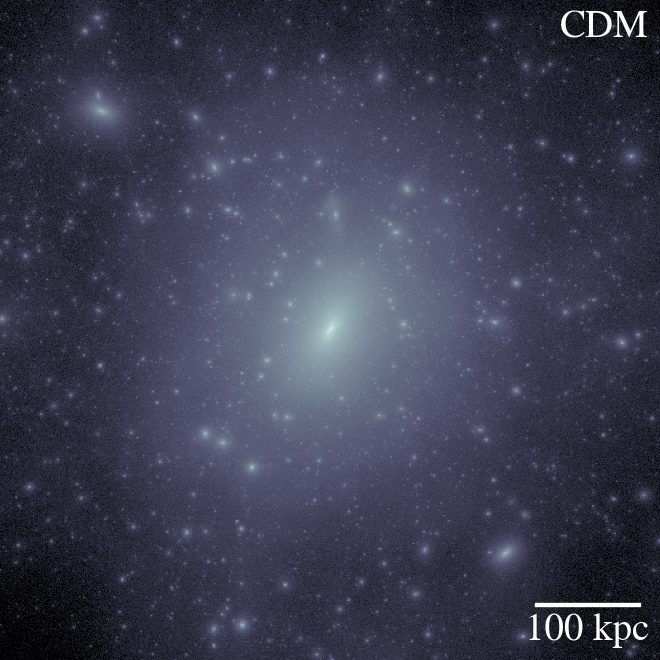}
    \includegraphics[width=0.32\textwidth]{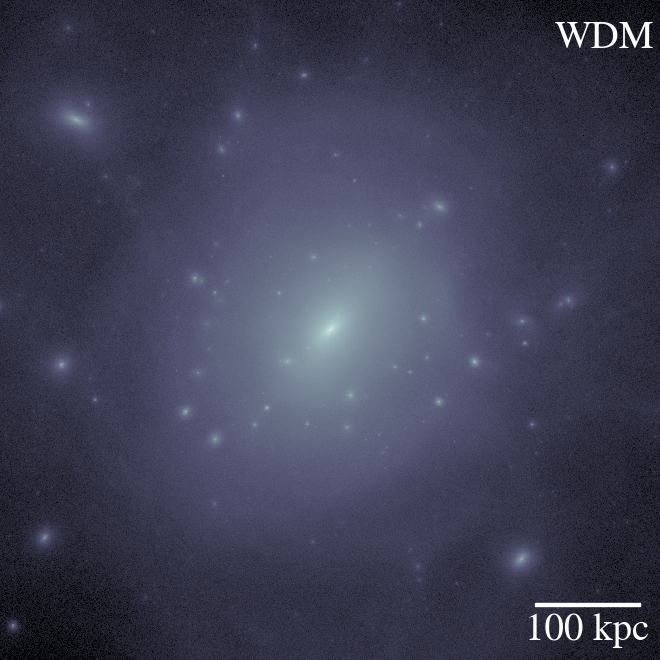}
    \includegraphics[width=0.32\textwidth]{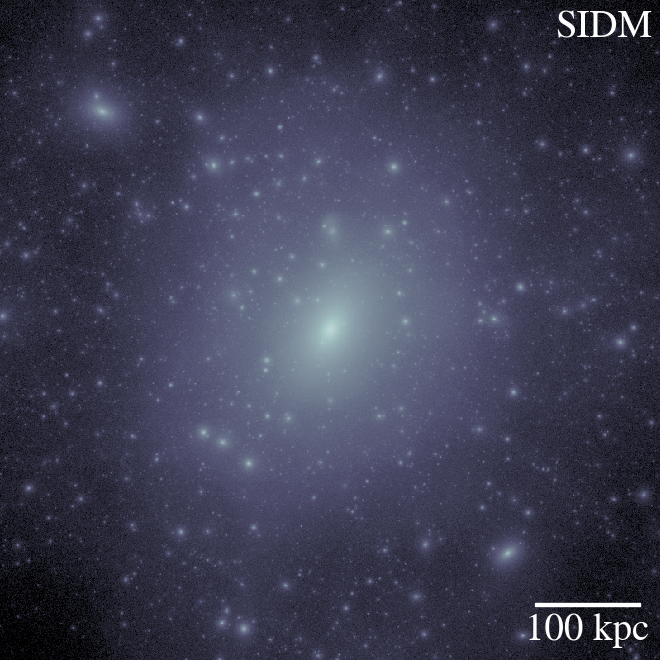}
    \caption{Cosmological simulations play a critical role in informing the observational effects of dark matter particle physics. This image demonstrates the simulated distribution of dark matter around a Milky-Way-mass galaxy in three different models of dark matter: CDM (left), a 2\,keV warm dark matter model (middle), and a self-interacting dark matter model with a cross section of $\sigma_{\rm SIDM}/m_\chi = 1~{\rm cm^2/g}$ (right). The effects of some models are immediately obvious by eye (e.g., the middle panel), while others can be detected at high statistical significance with cosmic observations (e.g., the right panel). Figure adapted from \citep{Bullock:2017xww}. \label{fig:cf3_bbk}}
\end{figure}

Cosmological N-body simulations are essential to predict and interpret the imprints of fundamental dark matter physics on structure formation in the nonlinear regime. Properly modelling baryon physics associated with galaxy formation in these simulations is often a key step to distinguish effects of baryon physics from those of new dark matter physics. With enormous datasets expected from Rubin LSST and other forthcoming facilities, we are at the critical stage to develop techniques for efficient forward simulations in order to extract information about dark matter from the data. In what follows, we give a brief overview and propose a plan for building simulation program to interpret observations so that we can robustly search for novel signatures of dark matter microphysics across a large dynamic range of length scales and cosmic time~\cite{Banerjee:2022qcb}.

Over the last $40$ years, cosmological simulations have played a vitally important role in studying dark matter particle properties. They have been essential to the development of the CDM paradigm and to eliminating neutrinos as a dominant component of the dark matter. During the last decade, particle physicists and simulators have come together to generate cosmological predictions for a subset of novel dark matter scenarios beyond CDM (e.g., Fig.~\ref{fig:cf3_bbk} \citep[][]{Lovell:2013ola,Vogelsberger:2015gpr,Du:2016zcv}). The challenge and opportunity for this decade is to develop a robust and vibrant simulation program that connects the ground-breaking capabilities of observational facilities \cite{Mao:2022fyx,Valluri:2022nrh,Dvorkin:2022bsc,Chakrabarti:2022cbu,Boddy:2022knd,Buckley:2017ijx} to an expanding landscape of particle models for dark matter and targeted terrestrial experiments \cite{Battaglieri:2017aum}. Because a well-synthesized program of theory, simulation, observation, and experiment is critical to revealing the nature of dark matter, we identify six areas of focus for simulations that advance along the key opportunities described in Sec.~\ref{sec:cf3summary}.

\begin{figure}[]
    \centering
    \includegraphics[width=0.75\textwidth]{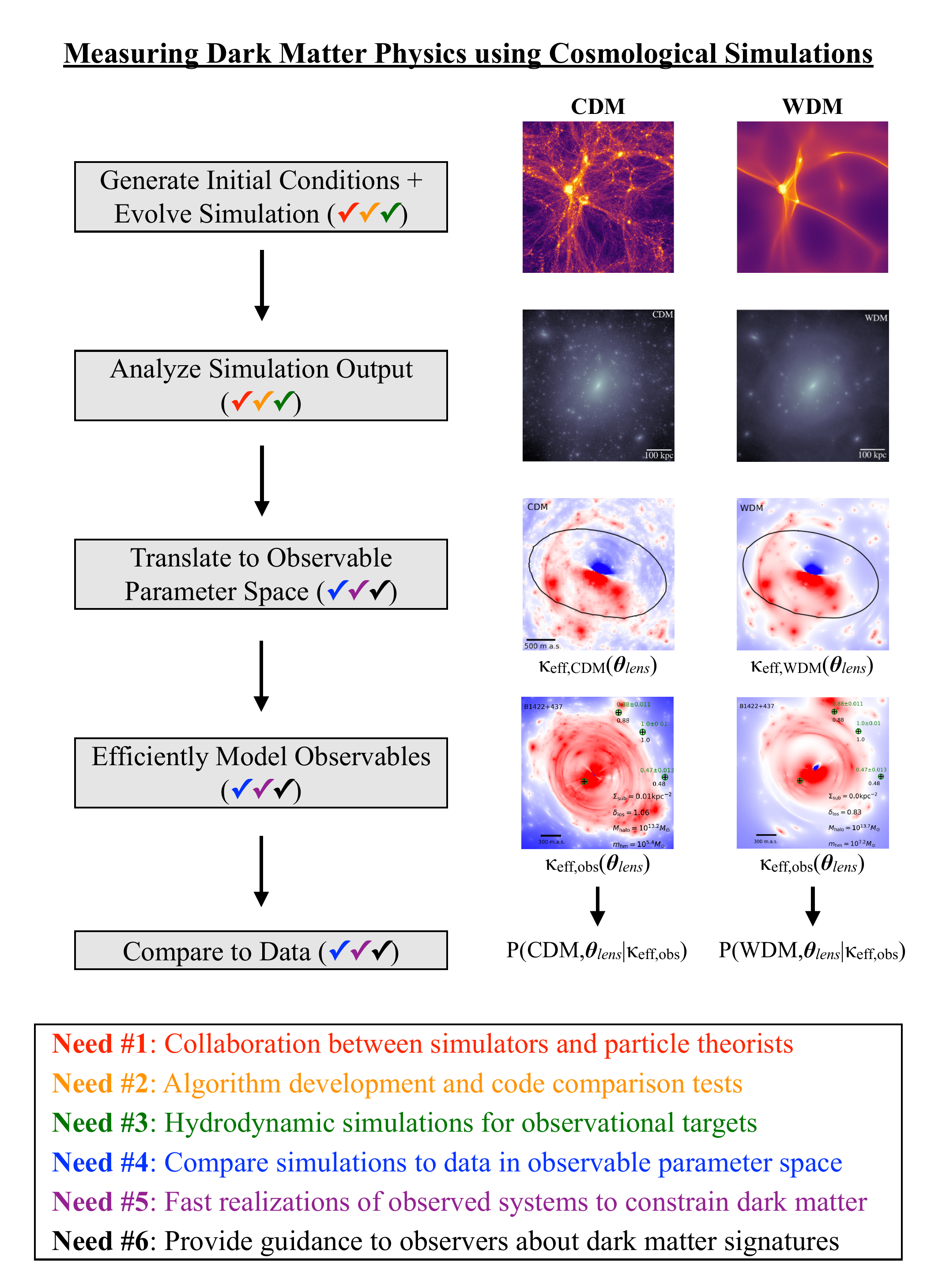}
    \caption{An example flowchart for distinguishing cold and warm dark matter models in the context of dark matter halo substructure as observed in strong gravitational lens systems. This example highlights the need for collaborative efforts among particle physicists, simulators and observers, in order to harness the full power of new observational facilities to quantitatively test dark matter models. The two right columns show images of simulations and lensing observables assuming cold and warm dark matter models. From top to bottom: large-box numerical simulations of structure formation, simulated dark matter substructure within a galaxy halo, a possible realization of dark matter structure generated under the model, and a particular realization of dark matter structure generated under the model consistent with observations of the strong lens system WGDJ0405-3308. Figure adapted from~\cite{Banerjee:2022qcb}.}
    \label{fig:cf03_sims}
\end{figure}

\noindent First, increased collaboration between simulators and particle theorists will help identify significant dark matter models and areas of parameter space for further study. Model builders and observers both rely on simulations as a crucial link that draws their ideas and work together. This approach underpins the key opportunity of using cosmic probes to understand fundamental properties of dark matter by mapping dark matter microphysics to astrophysical structure formation and observables associated with it. For example, knowing the scale on which structures are expected to be modified relative to CDM can enable simulators to efficiently target well-motivated regions of parameter space. In turn, targeted parameter space searches can help theorists focus their work on realistic model-building efforts. Guidance from theorists will be particularly valuable to rigorously develop initial conditions for simulations of specific dark matter models.

Second, it is important to advance algorithm development and develop code benchmarks to ensure that simulations meet the required precision targets set by the sensitivity of new facilities. Broadly speaking, there are four major classes of dark matter models that currently capture the attention of simulators: CDM, fuzzy dark matter, self-interacting dark matter, and warm dark matter. Each of these presents distinct challenges in numerical implementation, requiring benchmarks for validating simulations and ensuring that they achieve the necessary precision to successfully support dark matter inference. Key predictions include measurements of (sub)halo mass functions; (sub)halo density profiles; and subhalo radial distributions, infall times, and phase space distributions. 

Third, it is critical to perform simulations with full hydrodynamics using validated subgrid models and numerical resolution at the relevant redshifts and cosmological scales. Understanding the role of baryonic physics at small scales is critically important, since key discrepancies between the predictions of CDM and alternative dark matter models occur at small scales where baryonic physics plays an important role \citep{Bullock:2017xww}. Degeneracies between baryon physics and alternative dark matter models presents a challenge. Breaking these degeneracies requires full inclusion of baryonic physics in simulations and dedicated comparisons between validated simulations. Data from current and near-future facilities will usher in the discovery of many new types of systems with the potential to provide better sensitivity to dark matter physics, should support be provided for that specific scientific goal. 

Fourth, we will benefit significantly from the analysis of simulation outputs in the realm of observations. Forward modeling simulations into the space of observables enables apples-to-apples comparisons between models and data. Such investigations are necessary to fully prepare for and utilize unprecedented datasets from DESI, Rubin LSST, CMB-S4, and other forthcoming instruments. Rigorous comparisons between theory and observation, as well as tools that translate theoretical predictions to observable parameter spaces will continue to be essential. These comparisons will help us determine when a problem arises from numerical techniques and when it is a true physical problem. As data analysis pipelines and simulations become more elaborate -- and datasets become larger -- strengthening our capacity to disentangle numerical effects from physical phenomena will be of critical importance. Furthermore, translating simulations into observable parameter spaces will assist in designing and evaluating new facilities.

Fifth, we need fast realizations of observables to infer dark matter properties from observation on feasible timescales. 
Cosmological simulations with full hydrodynamics are a critical tool to reveal how the physical properties of dark matter alter the abundance and internal structure of dark matter halos and subhalos, which can result in observable differences in astronomical objects and systems. These simulations produce ``mock universes'' that allow us to compare theoretical prediction with observations in the space of observables. As such, running these simulations will become the bottleneck of parameter inference and model comparison, because these tasks typically require the generation of a large sample of simulated datasets covering different input parameters (dark matter properties in this case).
Multiple methods have been identified to address these challenges \cite{Banerjee:2022qcb}. 
They broadly fall into the categories of
(1) reducing the computational cost of individual simulations by swapping some simulation components with models, and (2) reducing the number of simulations needed for analyses. 
We will likely need to combine these approaches to cover the vast space of untested dark matter theories and the diversity of observational measurements. 
These efforts will benefit from the introduction of machine learning and artificial intelligence techniques that are described later in this chapter.

Finally, numerical simulations and fast realizations should inform observers and experimentalists where to look for new signatures of dark matter physics. Simulations can play a major role in motivating new observational strategies by revealing unforeseen signatures of, and affirming analytic predictions for, dark matter physics. One example of this dynamic in operation is the development of an accurate model for the dark matter distribution in our Milky Way galaxy. Understanding the local density and velocity distribution of dark matter is key to properly design terrestrial direct detection experiments. Furthermore, in the event that a positive signal is detected in a terrestrial experiment, we need to interpret that signal in an astrophysical context and confirm whether it is consistent with what cosmic constraints on dark matter. Another example is the identification of dark matter (sub)halos that do not contain baryons, as predicted by CDM. Simulations can inform observational strategies to characterize phenomena that are potentially entirely sourced by non-luminous objects.

 In Fig.~\ref{fig:cf03_sims}, we show an example flowchart for distinguishing cold and warm dark matter models using strong gravitational lens systems with the collaborative efforts from the six areas discussed above. With upcoming observational and experimental facilitates, the next decade will be transformative in the HEP community's ability to learn about dark matter in the sky and in the lab. Numerical simulations of structure formation are a bridge between theory and observation. A close collaboration among simulators, particle physicists, and observers is essential to interpret observational data, break degeneracies between baryon physics and dark matter physics, and reveal the microscopic nature of dark matter. Only with a vibrant and cohesive simulation program will we be able to leverage the full power and precision of cosmic probes of dark matter.

\section{Primordial Black Holes and the Early Universe}


\begin{figure}[t]
    \centering
    \includegraphics[width=0.9\textwidth]{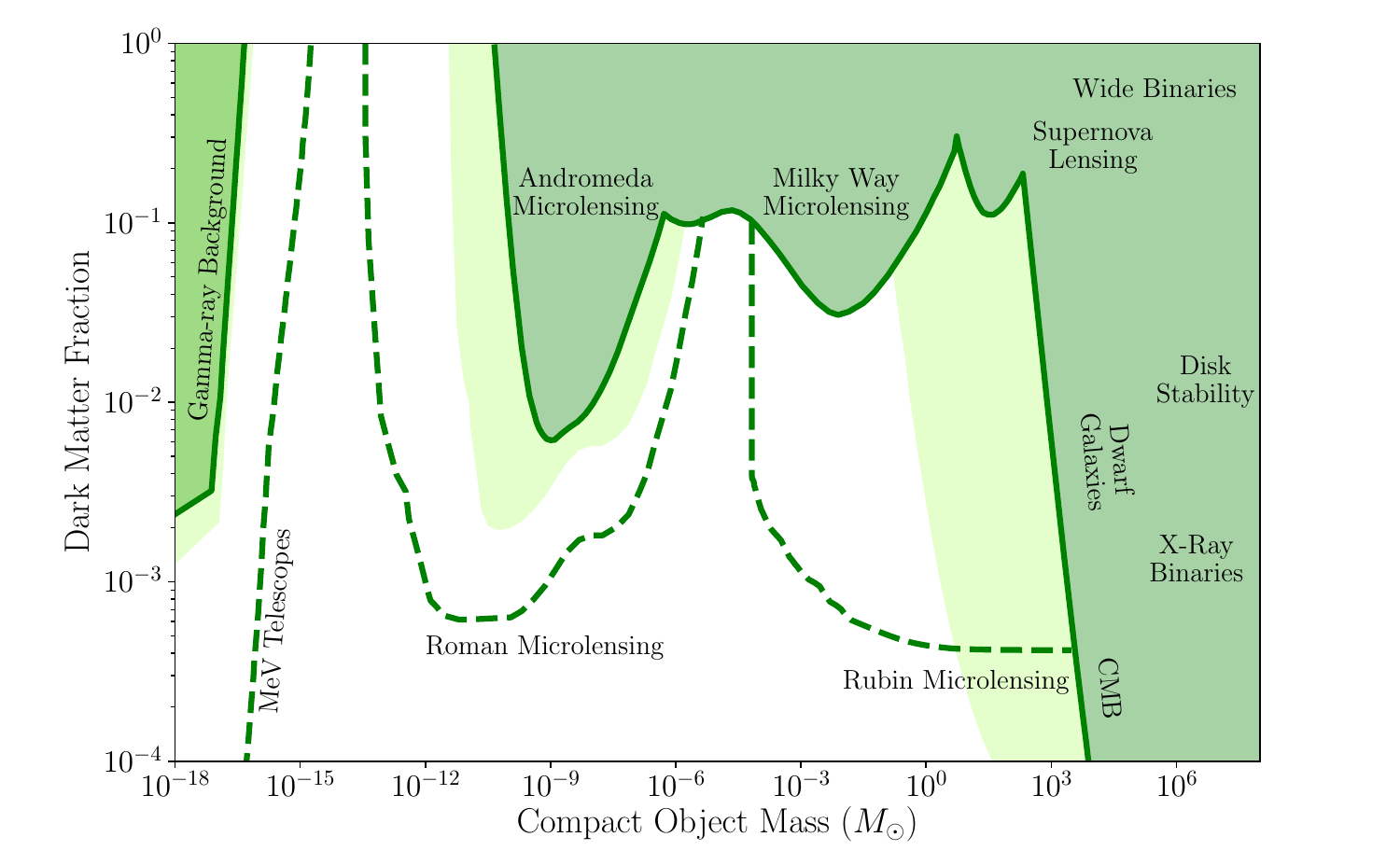}
    \caption{ \label{fig:cf3_pbh}
    Constraints on the dark matter fraction comprised of primordial black holes of a given mass. 
    Constraints derived with more and less conservative assumptions are shown in dark and light green, respectively.
    Projected sensitivity for future cosmic probes are shown with green dashed lines.  
    Existing constraints come from the gamma-ray background  \citep{Carr:2016drx}, microlensing observations of M31 with HSC ~\cite{Niikura:2019kqi,Smyth:2019whb}, microlensing observations of the Milky Way with MACHO/EROS \cite{MACHO_2001, Tisserand_2007, Wyrzykowski_2011}, supernovae lensing~\cite{Zumalacarregui:2017qqd}, dynamical heating of dwarf galaxies \cite{Brandt:2016aco,DES:2016vji,Lu:2020bmd,Takhistov:2021aqx,Takhistov:2021upb}, wide binary stars~\cite{Quinn:2009zg,Yoo:2003fr}, X-ray binaries~\cite{Inoue:2017csr}, CMB distortions from accreting plasma in early universe~\cite{Ali-Haimoud:2016mbv,Ricotti:2007au}, and disk stability constraints~\cite{Xu:1994vb}. 
    Dashed lines show the projected sensitivity of a Rubin LSST microlensing survey of the Galactic Bulge \citep{Drlica-Wagner:2019mwo}, a dedicated {\it Roman} microlensing survey of M31, and a future MeV gamma-ray facility \cite{Coogan:2020tuf,Ray:2021mxu}.
    Figure adapted from \citep{Bird:2022wvk,Drlica-Wagner:2019mwo}.}
\end{figure}

As potentially the first density perturbations to collapse, primordial black holes may be our earliest window into the birth of the universe and energies between the QCD phase transition and the Planck scale. The corresponding length scales ($k = 10^{7} - 10^{19}$ $h\,\mathrm{Mpc}^{-1}$) are much smaller than those measured by other current and future cosmological probes, see Fig.~\ref{fig:cf3_pk}. While previous estimates suggested that primordial black holes were constrained to be a subdominant component of dark matter over much of the viable mass range, more recent analyses have relaxed many of these constraints, re-opening the possibility that, in certain mass ranges, primordial black holes may comprise a dominant component of dark matter, as shown in Fig.~\ref{fig:cf3_pbh}.

The detection of primordial black holes would change our understanding of the fundamental physics of the early universe and the composition of dark matter \citep{Bird:2022wvk,Brito:2022lmd}. Primordial black holes are a probe of primordial density fluctuations in a range that is inaccessible to other techniques. These curvature fluctuations are imprinted on space-time hypersurfaces during inflation, at extremely high energies, beyond those currently accessible by terrestrial and cosmic accelerators. Our understanding of the universe at these high energies ($\gtrsim 10^{15}$\,GeV) comes predominantly from extrapolations of known physics at the electroweak scale. 
Measurements of the primordial density fluctuations via the abundance of primordial black holes would provide unique insights into physics at these very high energy scales.

This significant reward motivates the development of several complementary techniques that are sensitive to primordial black holes and subject to different astrophysical systematics, such as gravitational microlensing, gravitational wave detection, and gamma-ray signatures of black hole evaporation. 
In many cases, the science of primordial black holes can be performed by facilities that have well-motivated and multi-faceted scientific programs, e.g., optical/near-infrared time-domain imaging surveys, gravitational wave detectors, precision astrometry from radio interferometry, future MeV--TeV energy gamma-ray facilities. That said, realizing primordial black hole science from these facilities often requires specialized observing schemes, dedicated data analysis, and devoted theoretical studies. Therefore, it is important to closely integrate scientific efforts with enabling facilities across the scientific funding agencies (DOE, NASA, NSF).


Current and near-future observations can provide unprecedented sensitivity to the search for primordial black holes. However, it is necessary to ensure that these facilities acquire their data with a cadence and sky coverage that enables the  searches~\cite{Drlica-Wagner:2019mwo,Street:2018a,Street:2018b}. In addition, the sensitivity of the searches will be maximized by combining datasets from multiple observational facilities. Development of joint processing and analyses of Rubin LSST, the Nancy Grace Roman Space Telescope ({\it Roman}), and Euclid will maximize the opportunity to detect primordial black holes. Furthermore, current and future gravitational wave facilities will provide an unparalleled opportunity to detect primordial black holes directly through gravity. These facilities include both ground-based detectors, such as LIGO and Cosmic Explorer, and space-based detectors, such as LISA and AEDGE \citep[e.g.,][]{Ballmer:2022uxx}.
    
The scale of current and near-future datasets and the complexity of analyses benefit from collaborative scientific teams. These teams will develop the tools to perform rigorous and sensitive searches for primordial black holes in current and near-future observational data. The computational challenges presented by these searches are well-matched to the capabilities of HEP scientists and facilities. In addition, theoretical studies will help us better understand the production mechanisms, clustering, and spin properties of primordial black holes. These characteristics will inform the expected abundance of black hole microlensing and gravitational-wave events and systematics with cosmic surveys, as well as the connections to primordial physics in the early universe. Furthermore, improved simulations of the merger rate of primordial black holes and of specific accretion rates will help inform observational constraints.

\section{Dark Matter in Extreme Astrophysical Environments}


\begin{figure}[t!]
    \centering
    \includegraphics[width=0.9\textwidth]{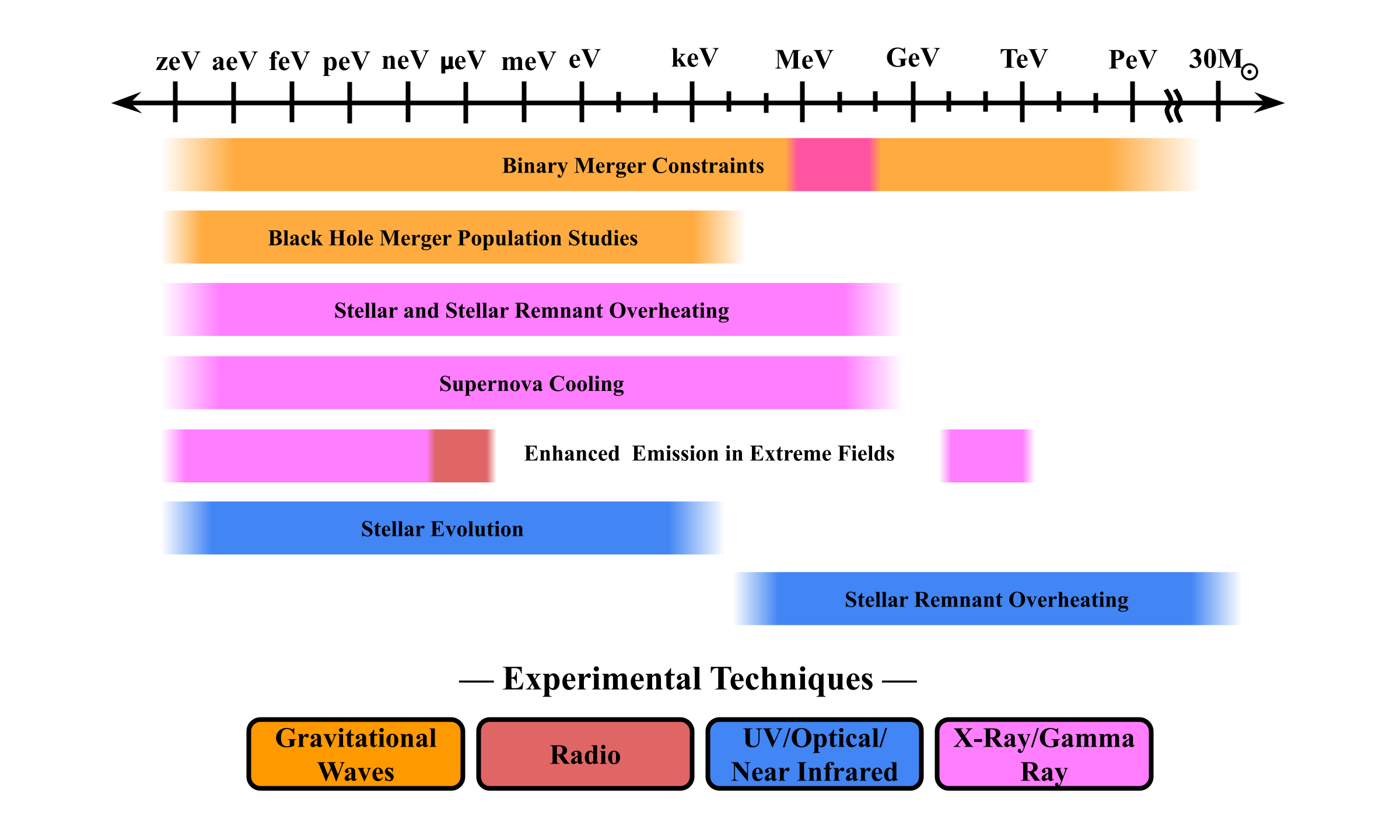}
    \caption{ \label{fig:extreme} A wide array of observations of extreme astrophysical environments have sensitivity to novel dark matter properties over ${\sim}50$ orders of magnitude in dark matter particle mass. Figure adapted from~\cite{Berti:2022rwn}.}
\end{figure}

Astro-particle searches for dark matter have historically focused on measuring cosmic-ray or photon products from the annihilation or decay of dark matter particles. However, dark matter interactions could also alter the physical processes occurring in the interiors of stars or stellar remnants, the dynamics of black holes, or the mergers of compact objects. These alterations would imprint characteristic signals in electromagnetic and gravitational wave observations. Exploring dark matter via observations of these extreme astrophysical environments---defined here as heavy compact objects such as white dwarfs, neutron stars, and black holes, as well as supernovae and compact object merger events---has been a major field of growth since the last Snowmass study. In the coming decade, observations of extreme astrophysical targets have the potential to open sensitivity to novel dark matter parameter space across a broad mass range (Fig.~\ref{fig:extreme})~\cite{Berti:2022rwn}. Exploiting these opportunities relies on both advances in theoretical work and on current and near-future observatories, including both gravitational-wave instruments and instruments spanning the full electromagnetic spectrum, from radio to gamma-rays. To help guide the discussion on the search in extreme astrophysical environments, we organize these searches by the dark matter mass range that is probed: ultralight dark matter ($<1$\,keV), light dark matter (keV--MeV), and heavy dark matter ($\gtrsim$\,GeV). Despite this categorization, we emphasize that many of these probes overlap in mass range, as summarized in Fig.~\ref{fig:extreme}. In addition, we note that the parameter space of the dark matter that is probed does not always saturate the relic abundance; instead, dark matter is broadly defined as matter that does not interact appreciably with Standard Model matter. 

Extreme astrophysical environments provide unique opportunities to probe ultralight dark matter ($<1$\,keV). Ultralight particles can be produced in the hot, dense cores of stars and stellar remnants and affect their evolution. Ultralight dark matter---either ambient in the environment or produced in a neutron star---can convert in the high magnetic field environment of the neutron star into radio waves or X-rays that could be detected by telescopes. In the last decade, new ideas unique to bosonic dark matter have been developed. Specific models of ultralight dark matter can alter the shape of the gravitational waveforms of merging neutron stars through their coupling to the dense neutron star matter. Black hole superradiance is a process that can extract energy and angular momentum from rotating black holes and place it into bound states of exponentially large numbers of ultralight bosons, as long as the Compton wavelength of the particle is comparable to the size of black holes. These systems yield signals of coherent gravitational waves as well as black hole spin down, which do not depend on particle interactions but only on gravity. Finally, ultralight dark matter can form collapsed structures like compact halos and boson stars that could be detected using gravitational waves, microlensing, or electromagnetic signals.

\begin{figure}[t!]
    \centering
    \includegraphics[width=0.8\textwidth]{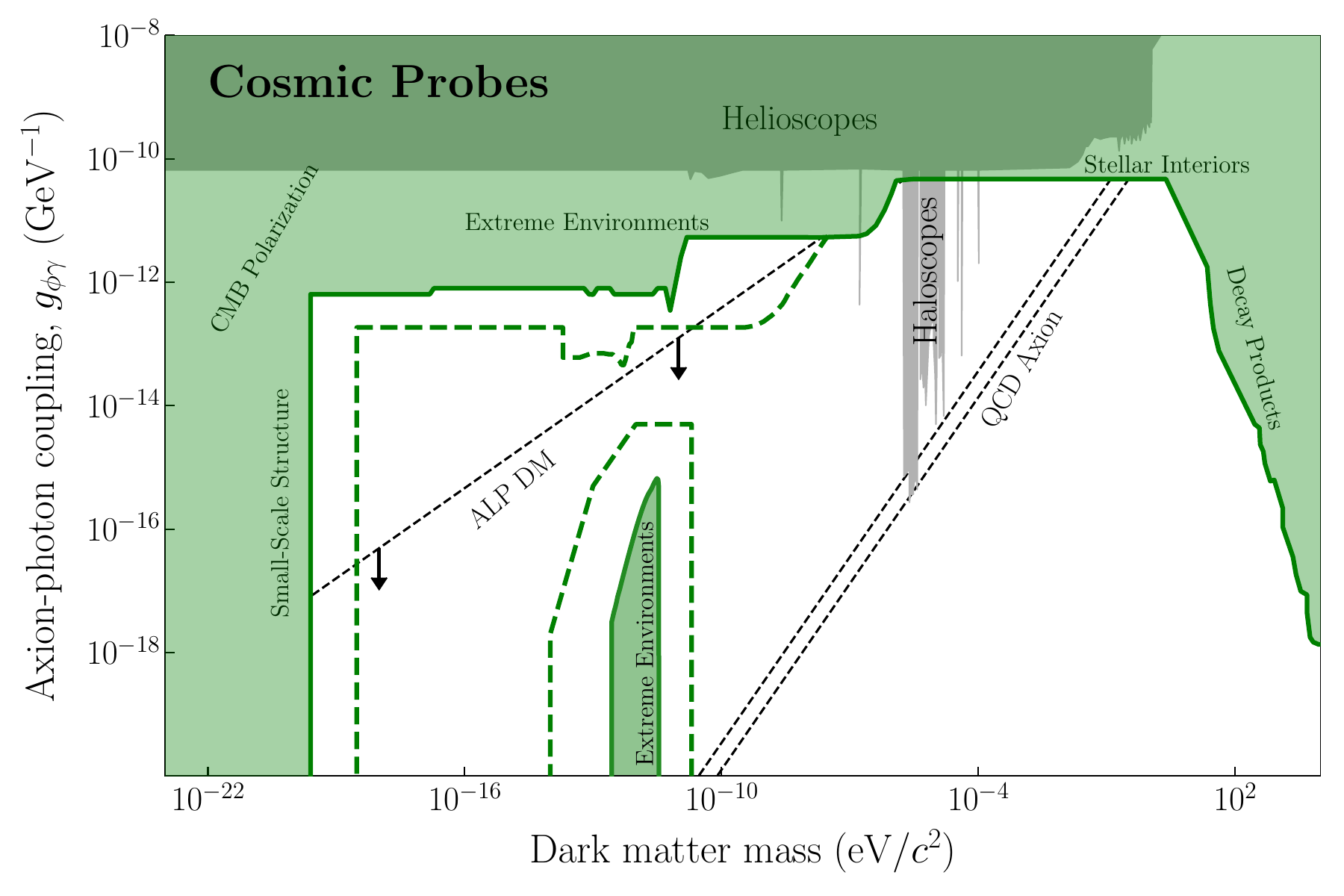}
    \caption{\label{fig:cf3_axion} Cosmic probes of extreme astrophysical environments \citep{Ayala:2014pea, Reynolds:2019uqt, Chen:2021lvo, Dessert:2022yqq, Baryakhtar:2020gao, Dolan:2022kul} combined with measurements of dark matter halos \citep[][]{Rogers:2020ltq, Nadler:2021dft} and other cosmological observations \citep[][]{Cadamuro:2011fd, Fedderke:2019ajk, Depta:2020wmr, Wadekar:2021qae} set strong constraints on the parameter space of axion-like particles (green regions). 
    Projected improvements in sensitivity coming from future facilities and observations are indicated with a dashed green line.
    Cosmic probes are sensitive well-motivated regions for the QCD axion \citep{Kim:1979if,Shifman:1978bx,Dine:1981rt,Zhitnitsky:1980tq} and axion-like particles \citep{Arias:2012az}, and they are highly complementary to other experimental searches with helioscopes and haloscopes (gray regions taken from \citep[][]{Adams:2022pbo}). This figure uses limit data available from \cite{AxionLimits}.}
\end{figure}

Opportunities to probe light dark matter (keV--MeV) exist from a variety of astrophysical situations: supernova explosions, the properties of neutron stars, binary neutron star mergers, and black hole population statistics (measured with gravitational waves from binary inspirals). Key observational targets for dark matter in this mass range include observation of gamma rays, neutrinos, and the populations of neutron stars and black holes as observed electromagnetically and via gravitational waves. Light dark matter produced in core collapse supernovae can be constrained from limits of their supernova cooling, or lead to visible signals in the X-ray or gamma-ray bands. During a binary neutron star merger, dark matter can lead to a bright transient gamma-ray signal. In the cores of blue supergiants, dark matter can affect stellar evolution, ultimately changing black hole population properties including the location of the black hole mass gap. Dark matter scattering and annihilating in exoplanets, brown dwarfs, Population III stars, and stellar remnants can be probed through infrared and optical radiation, and through gamma rays. Neutron stars can be heated by light dark matter via the Auger effect, which is probed by telescopes in the ultraviolet, optical and infrared ranges of the electromagnetic spectrum. Lastly, accumulation of dark matter (in particular bosonic light dark matter) can lead to the collapse of astrophysical objects. Most of the signals arise from couplings to Standard Model photons and fermions. As an example, in Fig.~\ref{fig:cf3_axion} we show the complementarity between cosmic probes of light and ultra-light dark matter and terrestrial helioscope and haloscope in the context of searches for the signatures of axion-like particles.

Compact astrophysical objects such as neutron stars and black holes provide unique environments to test heavy dark matter ($>$ GeV). Dark matter captured by neutron stars and their subsequent heating can be observed by upcoming infrared and radio telescopes. Dark matter that is produced in neutron stars may collect in the interior or form neutron star halos, with implications for the equation of state, mass-radius relation, and gravitational wave signals. Moreover, dark matter can collect in high-density spikes around black holes enhancing annihilation rates. A black hole--compact object binary can form a dark matter spike that can be observed by future space-based gravitational wave observatories. Merging compact objects can also give insight into a wide variety of dark sector particles that modify the dynamics of the merger process. This includes fifth forces and modifications to gravity. Finally, sufficient accumulation of dark matter around a compact object can cause the dark matter particles themselves to collapse into a black hole. Upcoming pulsar searches and gravitational wave observatories will be sensitive to this kind of dark matter signature. 

The coming decade presents a key opportunity to maximize the sensitivity of observations to novel dark matter theory space. Observational and theoretical astrophysicists should collaborate to constrain the standard astrophysical properties of these extreme environments. 
We need coordination between the experimental collaborations responsible for upcoming major observatories across all astrophysical messengers: gravitational waves, neutrinos and electromagnetic radiation at all wavelengths.
Theoretical developments will improve our understanding of the signatures of dark matter in extreme environments, which will in turn help optimize the design and data taking strategies of future instruments.

\section{Facilities for Cosmic Probes of Dark Matter}

Over the next decade, observational facilities spanning the electromagnetic spectrum, as well as gravitational waves, offer the potential to significantly expand our understanding of dark matter physics.
In this section, we briefly discuss current and near-future facilities that are aligned with the HEP Cosmic Frontier program and offer the opportunity to greatly enhance our understanding of dark matter physics.
In many cases, these facilities have multi-faceted scientific portfolios that include sensitivity to dark energy, inflation, neutrino physics, and modifications to gravity.
Furthermore, the technology used in these facilities leverages the core technical and scientific capabilities of the HEP community. 
Strong involvement from the HEP community will maximize the scientific output of these facilities, and in many cases, HEP involvement is necessary for the construction and/or operation of these facilities.
The capability to probe dark matter physics should be considered in the design phase of these new facilities.

The discussion in this section focuses on a series of facilities-oriented white papers submitted to CF3 as part of the Snowmass process \citep{Chakrabarti:2022cbu,Valluri:2022nrh,Mao:2022fyx,Dvorkin:2022bsc,Burns:2021pkx}.
We note that the facilities described here complement other multi-messenger facilities \citep{Engel:2022yig}, gamma-ray and X-ray experiments \citep{Aramaki:2022zpw,Engel:2022bgx}, and gravitational wave facilities \citep{Ballmer:2022uxx} submitted to other topical groups in the Snowmass process.

\subsection{Current/Near-Future Facilities}

\noindent {\bf Dark Energy Spectroscopic Instrument} \\ 
The Dark Energy Spectroscopic Instrument (DESI) began regular operations in 2021 and is currently performing one of the most powerful  wide-field spectroscopic surveys \citep{DES:2016vji,DESI:2019jxc}.
The synergy between DESI and other current and near-future observing facilities will yield datasets of unprecedented size and quality, with sensitivity to dark matter physics over a wide range of scales and redshifts. 
DESI will detect four times as many Lyman-$\alpha$ quasars as were observed in the largest previous survey, yielding about 1 million medium-resolution spectra.
Observations of these spectra will constrain the formation of dark matter halos through measurements of the clustering of low-density intergalactic gas out to $z \sim 5$. 
Measurements of stellar radial velocities from DESI in conjunction with astrometry from \Gaia (and eventually {\it Roman}) will enable us to constrain the global distribution of dark matter within the Milky Way, its dwarf satellites, and stellar streams. 
However, suites of numerical simulations of non-CDM cosmologies are needed to interpret observations from DESI in the context of fundamental dark matter physics. 
Such simulations must be transformed into realistic mock datasets that account for observational selection effects. 
The creation of these mock datasets is a significant investment that could heavily leverage HEP infrastructure that is already integrated into the DESI Collaboration~\cite{Valluri:2022nrh}.
The proposed DESI-II \citep{DESI2:2021,Schlegel:2022vrv} would extend the science capabilities of DESI through an extended operational period.

\noindent {\bf Vera C.\ Rubin Observatory} \\
The Rubin Observatory Legacy Survey of Space and Time (LSST), which is scheduled to start in 2024, has the potential to become a flagship dark matter experiment \citep{LSST:2008ijt}. 
LSST will probe dark matter through a wide suite of observations including measurements of Milky Way satellites, stellar streams, galaxy clusters, weak lensing, microlensing searches for primordial black holes, and studies of stellar populations and stellar remnants \citep{Drlica-Wagner:2019mwo,Bechtol:2019acd,Mao:2022fyx}.
Due to the size and complexity of the Rubin LSST dataset and the need for devoted, high-resolution numerical simulations, a coordinated effort is required to perform rigorous dark matter analyses.  
A large collaborative team of scientists with the necessary organizational and funding support is needed to lead this effort. Furthermore, studies of dark matter with Rubin LSST will also guide the design of, and confirm the results from, other dark matter experiments. Transforming Rubin LSST into a dark matter experiment is key to achieving the dark matter science goals that have been identified as a high priority by the HEP community~\cite{Mao:2022fyx}.

\noindent {\bf CMB-S4} \\
CMB-S4 is a ground-based experiment that will perform exquisite measurements of the CMB temperature and polarization anisotropies \citep{Abazajian:2019eic,CMB-S4:2022ght}. 
These measurements (on their own and in combination with other surveys) will provide new means to probe the nature of dark matter.
These measurements will provide a snapshot of the universe as it was around the time of recombination, and they will also reveal the imprints of structure growth at much later times. Gravitational lensing of the CMB leads to characteristic distortions of the primary CMB maps \citep{Lewis:2006fu}, allowing us to statistically reconstruct maps of the integrated line-of-sight density. Scattering of CMB photons in galaxy clusters (the Sunyaev-Zel’dovich effect) \citep{Sunyaev:1980nv,Sunyaev:1980vz} allows for the identification of the most massive bound structures in the universe out to very high redshifts. Cosmological measurements in general, and CMB measurements in particular, provide insights into dark matter physics that are complementary to direct, indirect, and collider searches for dark matter. Cosmological observables are impacted by the influence of dark matter on the entire cosmic history. Dark matter constraints derived from cosmology do not rely on assumptions about the dark matter density in the neighborhood of the Earth or of any specific astrophysical object. Furthermore, CMB observations are sensitive to regions of parameter space that are out of reach of current direct searches. 
Several aspects of the dark matter program are already included among the CMB-S4 core science cases; however, support must be provided to achieve these science goals~\cite{Dvorkin:2022bsc}.

\subsection{Future Facilities}

Here we briefly describe the landscape of proposed future facilities, starting with those probing the most energetic photons and moving to those with lower energies, and concluding with gravitational wave detectors. While these facilities are synergistic with a broad range of scientific objectives, strong support from the HEP community is necessary to enable the design, construction, and operation of these facilities. 

\noindent {\bf X-ray/Gamma-ray Facilities} \\
Instruments operating at X-ray and gamma-ray energies have been indispensable for the indirect dark matter detection and multi-messenger communities \citep{Aramaki:2022zpw,Engel:2022bgx,Engel:2022yig}. While indirect detection is discussed elsewhere, these experiments also provide an important test of PBH evaporation and dark matter in extreme astrophysical environments. 
In particular MeV-scale $\gamma$-ray experiments like AMEGO-X \citep{Caputo:2022xpx} and GECCO \citep{Orlando:2021get} are important for probing PBH evaporation, while X-ray experiments like XRISM \citep{Tashiro:2018}, Athena \citep{Barret:2018qft}, and proposed facilities such as STROBE-X \citep{STROBE-X:2019cyd} are important for probing the physics of extreme environments around neutron stars and black holes.

\noindent {\bf Optical/Near-Infrared Facilities} \\
Optical/near-infrared telescopes have been the work-horse for dark matter studies on galactic scales.
Proposed near-future optical/near-infrared facilities include a Stage-V Spectroscopic Facility (Spec-S5) \citep{Annis:2022xgg,Flaugher:2022rob} and the US Extremely Large Telescope program (US-ELT) \citep{US-ELT:2019}.
Both facility classes plan to target stars in the Milky Way halo, stellar streams, and dwarf galaxies, as well as dark matter-dominated galaxies in the local universe, strong lens systems, and galaxy clusters at higher redshift \citep{Chakrabarti:2022cbu}.
A Spec-S5 would combine a large telescope aperture, wide field of view, and high multiplexing, enabling it to obtain medium- to high-resolution spectra of millions of stars while simultaneously providing information on tens of millions of higher-redshift objects distributed over wide areas of sky \citep{Schlegel:2022lza}.
Proposed concepts for a Spec-S5 include MegaMapper \citep{Schlegel:2019eqc,Schlegel:2022vrv}, the Maunakea Spectroscopic Explorer (MSE) \citep{Li:2019nud}, and the ESO SpecTel concept \citep{Ellis:2019}.
In contrast, the US ELTs, including the Giant Magellan Telescope (GMT) \citep{GMT:2018} and the Thirty Meter Telescope (TMT) \citep{TMT:2015pvw}, will provide unprecedented sensitivity, image resolution, astrometry, and extreme precision radial velocity observations, but with much smaller fields of view and multiplexing than a Spec-S5.

In all cases, these facilities seek to use a variety of methods to extend measurements of the dark matter halo mass function below the threshold of galaxy formation ($10^6$--$10^8 M_\odot$) and to measure the density profiles of dark matter halos both in the local universe (e.g., for our Milky Way, its satellites, and other nearby galaxies) and at higher redshifts (e.g., strong lens systems and galaxy clusters).
Measurements of the dark matter halo mass function and density profiles can be translated into sensitivity to the dark matter particle mass and interaction cross-sections.
Additionally, these facilities provide multi-faceted fundamental physics programs that include measurements of dark energy and inflation \citep{Annis:2022xgg,Schlegel:2022lza}.
They would leverage HEP technology developed for Stage III and Stage IV dark energy experiments (DES, DESI, Rubin LSST) and advance technology toward a Stage V dark energy experiment \citep{Schlegel:2022lza}.  
HEP technical expertise is well-matched to the design and construction of fiber positioners, CCD detectors, and spectrographs. 
Furthermore, HEP computational resources and expertise are well matched to the task of data processing and distribution.
These experiments have strong support from the astronomy community; however, HEP support will be critical to enable their construction and to ensure that they maximize fundamental physics output.

\noindent {\bf Microwave Facilities}\\
The proposed millimeter-wavelength facility CMB-HD \citep{CMB-HD:2022bsz} will extend the resolution of cosmic microwave background surveys by a factor of five and the sensitivity by a factor of three or more.  These observations will open a new window of small-scale CMB observations and will uniquely enable measurements of the small-scale matter power spectrum (scales of $k \sim 10\,h\,{\rm Mpc}^{-1}$) from weak gravitational lensing using the CMB as a backlight.  These observations will also enable measurements that rule out or detect any new light particle species ($N_{\rm{eff}}$) that were in thermal equilibrium with the Standard Model at any time in the early Universe \citep{Dvorkin:2022jyg}, and enable probes of axion-like particles through CMB photon conversion, time-dependent CMB polarization rotation, cosmic birefringence, and ultra-high-resolution kinetic Sunyaev-Zel'dovich measurements.  CMB-HD would leverage and extend the scientific and technical investment of HEP in CMB-S4.

\noindent {\bf Radio Facilities} \\
Proposed centimeter-wavelength radio observatories, including the ngVLA \citep{Selina2018} and DSA-2000 \citep{Hallinan2019}, can employ pulsar timing measurements to map the dark matter halo of the Milky Way and the substructures it contains. These experiments could complement other gravitational wave facilities at higher frequency.
Proposed low-frequency radio experiments, such as LuSEE Night  \citep{Burns:2021pkx,Burns:2021}, PUMA \citep{PUMA:2019jwd}, and successors to HERA \citep{DeBoer:2016tnn} can use the 21-cm line of hydrogen from the Dark Ages ($z\sim50$) through cosmic dawn and reionization ($z \sim 6$) to probe dark matter physics via the thermal history of intergalactic gas and the timing of the formation of the first stars and galaxies.
These facilities would have complementary programs to probe dark energy (measuring the expansion history and growth of the universe up to $z=6$) and the physics of inflation (constraining primordial non-Gaussianity and primordial features).

\noindent {\bf Gravitational Wave Facilities} \\
Proposed gravitational wave facilities, such as Cosmic Explorer and LIGO-Voyager, can probe dark matter directly through gravity \citep{Ballmer:2022uxx}. These experiments are sensitive to channels including the detection of axion-like particles in binary neutron star mergers, ultralight bosons through superradiant instabilities of rotating black holes, the identification of boson stars in compact binaries, dark matter density spikes around black holes, and the existence of sub-solar-mass primordial black holes.

\section{Tools for Cosmic Probes of Dark Matter Physics}

\noindent {\bf Collaborative Infrastructure} \\
Historically, many cosmic probes of dark matter physics have been pursued by small groups of scientists. 
However, as the scale and complexity of cosmic survey experiments increase, the need for numerical simulations to interpret data grow, and the range of possible dark matter models expands, it becomes difficult to find sufficient expertise within a small group of scientists.
Similar challenges have been faced by the dark energy community, which has motivated the formation of large collaborative efforts to build and analyze data from new facilities.
These collaborations bring together the efforts of university groups, international collaborators, and scientists at national laboratories to accomplish scientific tasks that are too large for any single investigator.
Modern efforts to assemble collaborative teams to study cosmic probes of dark matter physics have already started in the context of the Dark Energy Survey~\citep{Nadler:2020fxi} and the Rubin LSST Dark Energy Science Collaboration \citep{Mao:2022fyx}.
In many cases, these teams can reside within existing collaborative infrastructure that has been established for other HEP mission goals.
However, additional support must be provided to enable  dark matter as a parallel branch of fundamental physics being pursued by these experiments.

\noindent {\bf New Support Mechanisms}\\
Cosmic probes of dark matter provide a rich, diverse, and interdisciplinary area of research.
While this leads to an exciting discovery space, it also leads to logistical difficulties in classifying the research within the existing research support structures (i.e., DOE HEP, NSF-PHY, NSF-AST, and NASA).
In particular, support for cosmic probes of fundamental dark matter properties often falls in the cracks between these disciplines and agencies.
This has been especially challenging for theoretical research in this domain, which has been increasingly been difficult to fund.
We recommend that inter-agency coordination is required to assure that cosmic probes of dark matter physics are firmly supported.
Multi-agency support extending across the spectrum of theory, simulation, and experiment will enable large gains in the coming decade.

\noindent {\bf Artificial Intelligence/Machine Learning}\\
The interplay between models and observations is a cornerstone of the scientific method, aiming to inform which theoretical models are reflected in the observed data. Within cosmology, as both models and observations have substantially increased in complexity over time, the tools needed to enable a rigorous comparison have required updating as well. In the next decade, vast data volumes will be delivered by ongoing and upcoming cosmology experiments, as well as the ever-expanding theoretical search space. We are now at a crucial juncture where we may be limited by the statistical and data-driven tools themselves rather than the quality or volume of the available data. Methods based on artificial intelligence and machine learning have recently emerged as promising tools for cosmological applications, demonstrating the ability to overcome some of the computational bottlenecks associated with traditional statistical techniques. Machine learning is starting to see increased adoption across different sub-fields of and for various applications within cosmology. At the same time, the nascent and emergent nature of practical artificial intelligence motivates careful continued development and significant care when it comes to their application in the sciences, as well as cognizance of their potential for broader societal impact \citep{Dvorkin:2022pwo}. 

\noindent {\bf Cosmology Data Preservation} \\
Cosmology datasets and simulations have useful lifetimes that extend long beyond the operational period of individual projects.
As datasets from facilities and simulations grow in size, the ``take out'' model of manual download followed by local computation will become insufficient and unwieldy.
Future work needs to focus on co-locating data with computing, and automating the coordination
between multiple data/compute centers.
Furthermore, special attention should be paid toward facilitating the joint analysis of datasets beyond the lifetime of individual projects. 
Implementing a comprehensive data preservation system will require support not only for hardware, but also for personnel to develop and maintain the technologies to simplify cross-site data sharing and personnel to curate the relevant datasets.
The authors of \citep{Alvarez:2022kut} recommend that the HEP community support a new cosmology data archive center to coordinate this work across multiple HEP computing facilities.

\section{Roads to New Physics}

Cosmic probes of dark matter present exciting opportunities with powerful synergies to other dark matter search techniques, namely ground-based direct and indirect detection facilities. 
In this section, we consider three exemplar scenarios where astrophysical observations lead to the characterization of fundamental dark matter properties, and discuss their implications for revealing the particle nature of dark matter and understanding early-universe cosmology. In particular, we  highlight how astrophysical observations, terrestrial experiments, and theoretical modelling can work together to extract fundamental parameters of dark matter. 

\subsection{Warm and Self-Interacting Dark Matter}

We first consider a scenario where dark matter differs from the standard CDM model by having a warm thermal velocity distribution and an appreciable self-interaction cross section. The matter power spectrum of warm dark matter is suppressed compared to CDM, resulting in a reduction in the number and the central density of low-mass dark matter halos. For thermal warm dark matter, the power spectrum is completely determined by the dark matter particle mass, $m_{\rm WDM}$ \citep{Bullock:2017xww}. The observed population of satellite dwarf galaxies of the Milky Way can be used to measure $m_{\rm WDM}$ if such a suppression is observed. Furthermore, dark matter self-interactions can thermalize the inner regions of dark matter halos and change the dark matter density profile. Thus, the self-interaction cross section per unit mass, $\sigma_{\rm SIDM}/m_\chi$, can be inferred from measurements of stellar velocities of galaxies, which are sensitive to the central dark matter content. 

 \begin{figure}[t]
    \centering
    \includegraphics[width=0.8\textwidth]{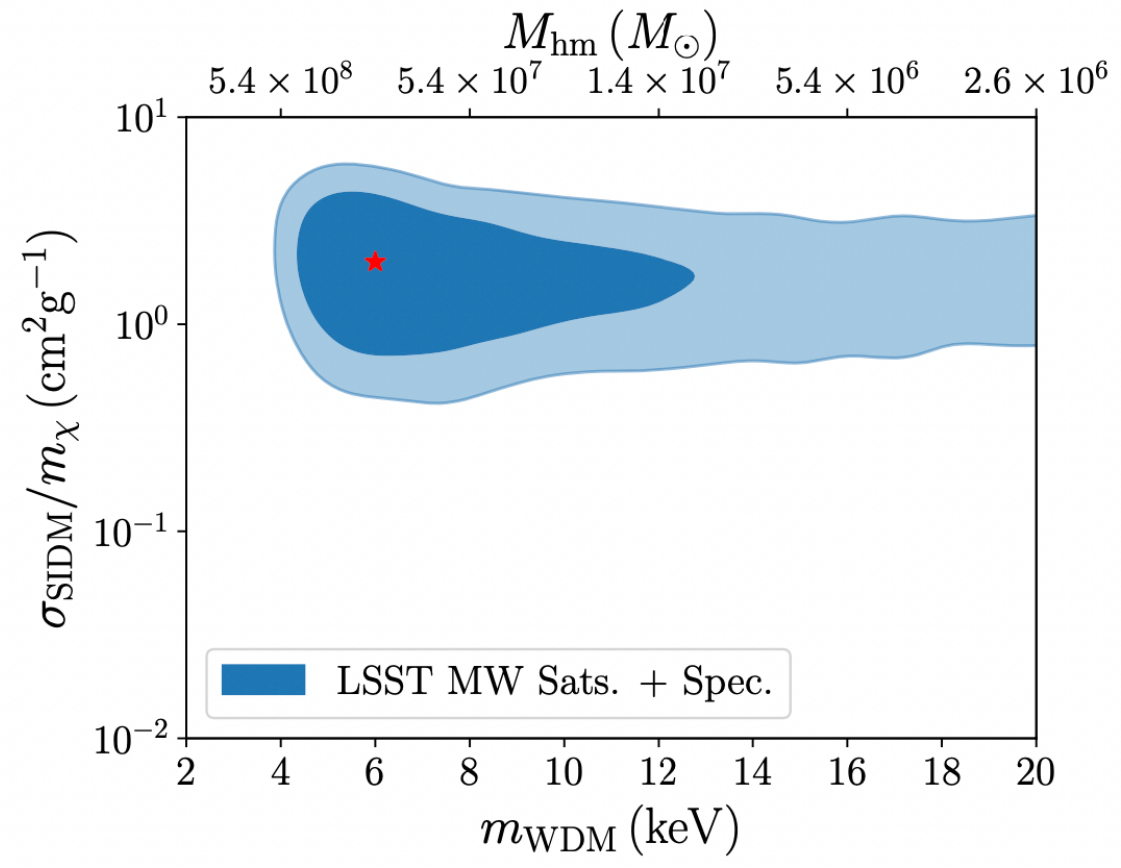}
    \caption{\label{fig:cf3_sidmwdm} Potential measurements of the self-interacting cross section and warm dark matter mass from upcoming observations by the Rubin Observatory and a future Stage-V Spectroscopic Facility (e.g., MSE \citep{Li:2019nud} or MegaMapper \citep{Schlegel:2019eqc,Schlegel:2022vrv}). The projection assumes a dark matter model with a cross section of $\sigma_{\rm SIDM}/m_\chi=2~{\rm cm^2/g}$ and a suppressed matter power spectrum corresponding to a warm dark matter mass of $m_{\rm WDM} = 6~{\rm keV}$ (red asterisk). The uncertainties contours are created by following a procedure similar to~\cite{Nadler:2018iux}. Figure adapted from~\cite{Drlica-Wagner:2019mwo}.
    }
\end{figure}

Fig.~\ref{fig:cf3_sidmwdm} demonstrates the ability of Rubin LSST combined with a future spectroscopic survey to measure the particle properties of dark matter from observations of Milky Way satellite galaxies, adapted from~\citep{Drlica-Wagner:2019mwo}. This projection assumes a self-interaction cross section of $\sigma_{\rm SIDM}/m_\chi=2~{\rm cm^2/g}$ and a matter power spectrum corresponding to thermal warm dark matter with $m_{\rm WDM}=6~{\rm keV}$; see ~\cite{Drlica-Wagner:2019mwo} for details.\footnote {More recent studies show that a larger cross section on dwarf scales may be needed to fully explain diverse dark matter densities of the Milky Way satellite galaxies in self-interacting dark matter models~\cite{Nishikawa:2019lsc,Kaplinghat:2019svz,Sameie:2019zfo,Zavala:2019sjk,Correa:2020qam,Turner:2020vlf,Silverman:2022bhs}. In this case, dark matter (sub)halos with a high concentration can experience gravothermal collapse, resulting in a high inner density, which can be probed using strong lensing observations with Rubin LSST, see, e.g.,~\cite{Minor:2020hic,Yang:2021kdf,Gilman:2021sdr,Gilman:2022ida,Sengul:2022edu} for relevant discussion. } The projection should be regarded as an illustration of the capability of future facilities to \emph{measure} fundamental dark matter particle properties using observations of the cosmic distribution of dark matter. This measurement does {\it not} assume that dark matter couples to the Standard Model. Furthermore, we can break degeneracies between dark matter particle properties and the physics of galaxy formation (e.g., the long tail towards large dark matter mass) by combining satellite galaxy measurements with a probe that is independent of subhalo luminosity, such as strong lensing and stellar streams, ultimately resulting in closed contours at high statistical significance.

Astrophysical observations can provide precision measurements of interactions in the dark sector. To achieve this goal, a collaborative effort is crucial. First, after Rubin LSST discovers new satellite galaxies, spectroscopic followup measurements of their stellar velocity dispersion are needed to constrain the dark matter density. Second, with the population of newly-discovered satellites, it will be possible to update models that capture the connection between invisible subhalos and visible galaxies to better control baryonic uncertainties. Third, dedicated N-body simulations are needed to make concrete predictions of self-interacting and warm dark matter models in terms of the properties of subhalos, such as their mass function, orbital parameters and central density. Fourth, in order to implement the novel dark matter properties in the simulations and interpret the observational results, we need to use the methods of particle physics to calculate the self-interaction cross section and determine how it depends on the velocity and scattering angle, as well as the linear matter power spectrum that encodes the evolution of the dark matter candidate in the early universe. Last but not the least, we will make predictions for observables on different galactic systems, such as density profiles in isolated dwarf galaxies and substructures of galaxy clusters, and search for signatures of new physics beyond CDM in these systems.

The possibility that dark matter has strong self-interactions indicates that there is a light mass scale ${\cal O}(1)~{\rm MeV}$ in the dark sector, which is much below the weak scale. Such a light dark sector motivates dedicated searches in upcoming and proposed terrestrial experiments in the intensity frontier, such as FASER~\cite{Feng:2022inv} and LDMX~\cite{Markiewicz:2022xis}; see~\cite{Ilten:2022lfq} for an overview of relevant experiments and facilities. It is also natural to expect that the light dark sector contains relativistic degrees of freedom~\cite{Cyr-Racine:2015ihg,Huo:2017vef}, i.e., dark radiation, which can be probed in CMB-S4~\cite{CMB-S4:2016ple}. In addition, after combining observations from dwarf to cluster scales, spanning from $\sim10^8~M_\odot$ to $10^{15}~M_\odot$, we may determine masses of dark matter and mediator particles~\cite{Kaplinghat:2015aga}. The discovery of a large self-interaction cross section and a cut-off in the matter power spectrum will have profound implications for constructing particle theories of dark matter and understanding its evolutionary history in the early universe, see~\cite{Tulin:2017ara,Cyr-Racine:2015ihg,Buckley:2017ijx,Adhikari:2022sbh} for related discussion.

\subsection{Primordial Black Holes}

\begin{figure*}[t]
    \centering
    \includegraphics[width=0.8\textwidth]{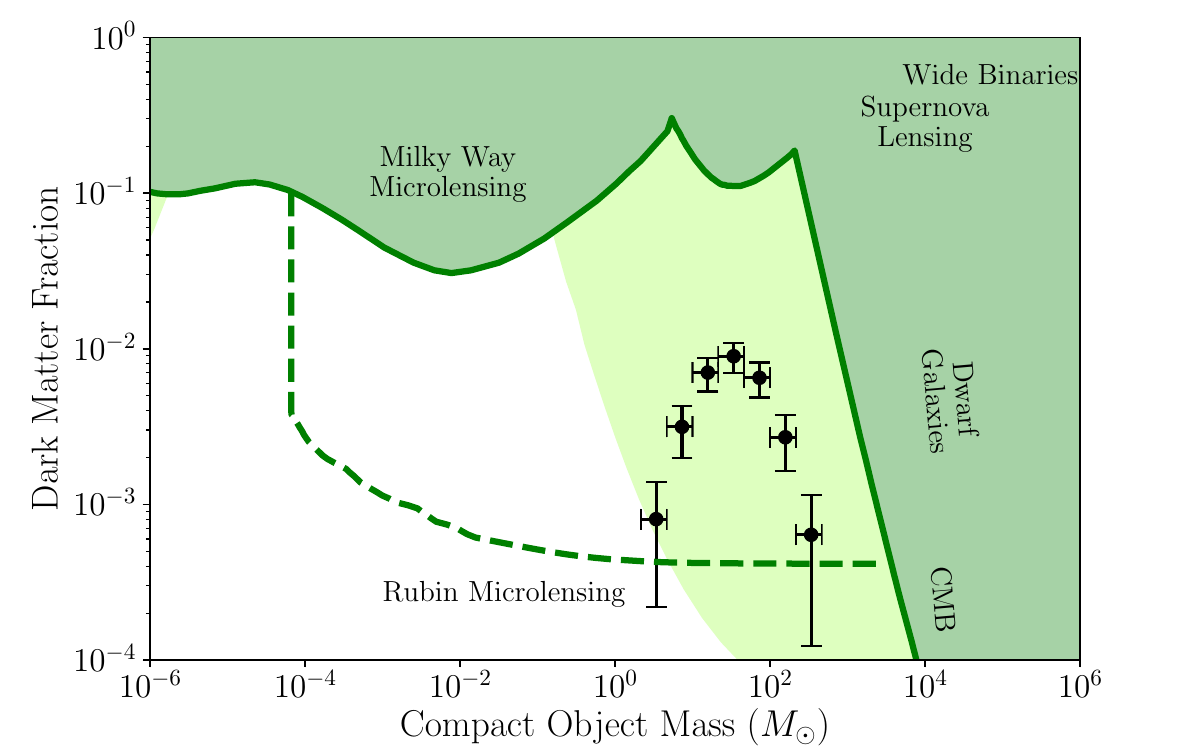}
 \caption{\label{fig:cf3_phbdiscovery} Projected sensitivity of Rubin LSST to primordial black holes that make up a subdominant fraction of the dark matter (black circles). This figure assumes a log-normal mass function that provides an integrated contribution of 3\% of the dark matter \citep{Carr:2017jsz}. Fractional uncertainties on the projected measurements are estimated from the Poisson uncertainties on the predicted number of microlensing events that would be observed by Rubin LSST. Current (green regions) and future (dashed green) constraints on compact object dark matter come from Fig.~\ref{fig:cf3_pbh}.}
\end{figure*}

Next, we consider the discovery of primordial black holes, a realization of compact object dark matter. Primordial black holes represent a gravitational dark matter candidate that cannot be produced in terrestrial experiments and can only be detected and studied observationally. Much of the primordial black hole dark matter parameter space has been constrained by existing observations \citep[e.g.,][]{Carr:2016drx}, but windows still remain where primordial black holes can make up some or all of the dark matter, see Fig.~\ref{fig:cf3_pbh}. Even if primordial black holes make up a subdominant component of the dark matter, their existence would revolutionize our understanding of early universe physics at extreme temperatures that are inaccessible to laboratory experiments \citep{Bird:2022wvk}.

Rubin LSST provides an exciting opportunity to directly measure the mass function of compact objects through microlensing observations. 
If scheduled optimally, LSST will provide sensitivity to microlensing event rates corresponding to $10^{-4}$ of the dark matter density in compact objects with masses $\gtrsim0.1\,{\rm M_\odot}$, a factor $10^{2}\textup{--}10^{3}$ improvement compared to the existing limits. In addition, the {\it Roman Space Telescope} also provides microlensing opportunities in the search for primordial black holes in the next decade. As a high resolution space-based imaging system, {\it Roman} has the potential to detect or constrain primordial black holes through various types of lensing.

Fig.~\ref{fig:cf3_phbdiscovery} shows discovery potential for primordial black holes making up a subdominant component of the dark matter. We follow the prescription of Carr et al. 2017 \citep{Carr:2017jsz} and model the normalized primordial black hole mass function with a log-normal distribution (see their Eq. 3). 
We set the parameters of our mass function to be consistent with existing observational constraints \citep{Carr:2017jsz}, choosing a peak mass of $M_c = 30\,{\rm M_\odot}$, a width of $\sigma = 1$, and an integrated contribution to the dark matter density of $f_{\rm PBH} = 0.03$ (3\%).
We bin into logarithmic mass bins with width of $0.33$ dex and calculated the integrated contribution to the dark matter abundance and the expected number of microlensing events that would be observed by LSST using the projected sensitivity \citep{Drlica-Wagner:2019mwo}.\footnote{This width of our mass bins is roughly comparable to the mass uncertainty reported for the recent detection of a microlensing event \cite{Lam:2022vuq}.} 
We assign uncertainties on the measured compact object fraction from the fractional Poisson uncertainties on the number of observed events.
Rubin LSST will be sensitive to the existence of this primordial black hole population with high confidence; however, this analysis does not include contamination from astrophysical black holes, which is expected to be small at these high masses \citep{Lam:2019lza, Drlica-Wagner:2019mwo}. 

The detection of primordial black holes by Rubin LSST and/or {\it Roman} would provide insights into early universe cosmology. Primordial black holes could form at early times from the direct gravitational collapse of large density perturbations that originated during inflation. The same fluctuations that initialize seeds of galaxies, if boosted on small scales, can lead to some small areas having a Schwarzschild mass within the horizon, which collapse to form black holes. Thus the detection of primordial black holes would directly constrain the amplitude of density fluctuations. These constraints probe small scales between $k = 10^7\textup{--}10^{19}~{\rm h/Mpc}$, much smaller than those measured by other current and future probes. 

\subsection{Axions and Axion-like Particles}

Interest in axion and axion-like particles as potential dark matter candidates has increased significantly since the last Snowmass study \citep{Kusenko:2013saa,Adams:2022pbo,Antypas:2022asj}.
The diversity of terrestrial axion experiments has increased dramatically during this time period \citep{Adams:2022pbo}. 
Current experiments now probe the QCD axion model space in the $\mu$eV mass range \citep{ADMX:2019uok,HAYSTAC:2020kwv,ADMX:2021nhd}, and future experiments are poised to extend sensitivity to higher and lower masses \citep{Adams:2022pbo}.
Meanwhile, the expanding theoretical landscape of axion-like particle models makes it increasingly important to understand the cosmological implications of these models and their detectable signatures in the cosmos.
For example, a thermal population of sub-eV mass axions would leave a cosmological imprint that could be detected through measurements of $N_{\rm eff}$ \cite{Dvorkin:2022jyg}.

Cosmic probes provide important complementary information in the search for QCD axions and axion-like particles. 
In particular, a detailed understanding of the local dark matter distribution is important for both designing terrestrial search strategies and interpreting any positive experimental results from those searches \citep{Green:2017odb}.
For example, many terrestrial cavity experiments search for a distinct, narrow feature coming from axion conversion into photons \citep{Adams:2022pbo}. 
The width of the axion feature depends on the velocity dispersion of dark matter in the Milky Way, and more precisely, the Solar neighborhood. Thus, precision cosmic measurements of the local dark matter velocity and density distributions are an important guide when planning future searches for axions and axion-like particles.  

The existence of dynamically cold dark matter substructure within the Solar System can change the signal strength and temporal modulation behavior in dark matter direct detection~\citep{Freese:2003na,Savage:2006qr}. In addition, gravitational focusing of dark matter streams by the Sun, Moon and Earth can affect the phase-space distribution of local dark matter particles \cite{Alenazi:2006wu,Sofue:2020mda,Lee:2013wza} and have an impact on direct detection signals accordingly. Many current studies are based on WIMPs models, and more work is needed to extend them to axion detection. For instance, since axions are much colder than WIMPs in the streams, the axion signal strength can be boosted more significantly due to gravitational focusing~\cite{focusing}. In fact, the possible existence of cold dark matter structures has motivated alternative ``high-resolution'' experimental scan strategies \citep{ADMX:2006kgb}. Conversely, the discovery of an axion signal in terrestrial haloscope experiment would immediately provide information about the local velocity dispersion of dark matter and enable improved modeling of the local dark matter distribution \citep{Foster:2017hbq}.

Cosmic probes can also help guide future terrestrial searches over the much broader parameter space of axion-like particles.
Cosmic observations are currently the most sensitive probes of axion-like particles coupling to photons in the mass range below $\roughly 10^{-6}$ eV, see Fig.~\ref{fig:cf3_axion}. 
In this regime, upcoming probes of extreme astrophysical environments, such as observations of neutron stars or a Galactic supernova, may provide positive signal of axion-like particles \citep{Berti:2022rwn}.
Such a signal could help guide the design of future terrestrial searches using lumped element approaches or superconducting radiofrequency cavities \citep{Adams:2022pbo}. 
While these terrestrial experiments are currently envisioned to cover large regions axion-like particle parameter space \citep{Adams:2022pbo}, their design and operation could be accelerated if cosmic probes identified a specific target mass range and coupling strength.
In the regime below $\roughly 10^{-12}$ eV, cosmic probes such as searches for axion-like particle condensates through microlensing~\citep{2021arXiv210803063S} and black hole superradiance \citep{Baryakhtar:2020gao} may yield a positive signal that could motivate the design of novel atomic clock or nuclear magnetic resonance experiments \citep{Antypas:2022asj}. Moreover, below $\roughly 10^{-12}$ eV, the length scales associated with equilibrium states for axion dark matter approach scales of astrophysical significance, making observations of cosmic phenomena an essential probe in this mass regime. For example, an axion-like particle with mass $\roughly 10^{-21}$ eV produces a dwarf galaxy-scale halo with a distinct soliton core~\cite{Schive:2014dra,Mocz:2017wlg,2022PhRvD.105l3540G}. Cosmic probes are currently our only option for experimentally testing models in this mass regime.

\section{Summary and Outlook}

More than 80 years after the first astrophysical discovery of dark matter, its fundamental nature remains an open question. Over the last several decades, the HEP community has designed and executed extensive searches for dark matter in a wide variety of terrestrial experiments. Despite these heroic efforts, the only positive measurements of dark matter continue to come from cosmic observations. Scientific advances over the last decade have made it possible to use precision measurements of macroscopic astrophysical systems to probe the microscopic particle properties of dark matter. This Snowmass report presents the critical opportunity for the HEP community to  fully realize the potential of cosmic probes of dark matter.

In this report, we have described methods of measuring fundamental properties of dark matter that are valid even when the coupling between dark matter and normal matter is extremely weak (e.g., as weak as gravity). Cosmic measurements of the distribution of dark matter, including the matter power spectrum, dark matter halos, and compact objects, can constrain particle properties of dark matter, such as the mass, interaction cross section, and production mechanism. Moreover, if dark matter has feeble non-gravitational interactions with normal matter, extreme astrophysical environments, such as neutron stars and black holes, provide unique opportunities to explore dark matter physics over 50 orders of magnitude in the mass; much of this model space is inaccessible with terrestrial experiments. In addition, precision astrophysical measurements of dark matter with current and near-future observational facilities are critical for interpreting results from conventional dark matter searches. 

We have further demonstrated that with the unprecedented coverage and sensitivity of current and near-future observational facilities, the rapidly improving scale and accuracy of numerical simulations, and better theoretical modelling, astrophysical uncertainties can be controlled and the fundamental parameters of dark matter can be extracted. This makes it possible to map Lagrangian parameters describing a particular dark matter model to astrophysical observables, and vice versa. Thus cosmic probes can provide precision measurements of particle physics properties of dark matter in a way that is similar to how HEP experiments have enabled the construction of the Standard Model of particle physics.

Cosmic probes of particle properties of dark matter have emerged as a new research field since the last Snowmass community study, largely due to tremendous progress in theoretical investigations of novel dark matter scenarios, N-body simulations of structure formation, as well as astrophysical observations of dark matter distributions. There is a new and exciting trend in the HEP community that more and more theoretical particle physicists have begun working on astrophysical aspects of dark matter. At the same time,  astrophysicists working on N-body simulations have started to develop simulation algorithms that can model novel dark matter scenarios beyond CDM. We must encourage and support this promising and evolving trend from both communities. 

Furthermore, we must develop new mechanisms to further support synergistic efforts among theorists, simulators, dynamicists, observers, and experimentalists/instrumentalists, who are traditionally supported by different agencies and/or programs. Cosmic probes of dark matter are fundamentally multidisciplinary and interdisciplinary, and  traditional disciplinary divisions limit scientific outcomes. New support mechanisms can be pursued from small to large scales. On small scales, a program like the DOE Dark Matter New Initiatives (DMNI) is well suited to support a small-scale collaborative effort from particle physicists and astrophysicists with a well-defined scientific goal. Cosmic probes of dark matter were not included in the current DMNI program. If the program continues, we strongly urge that DOE integrates cosmic probes into its portfolio. Alternatively, a similar program could be established to make rapid progress in this emerging field.

Dark matter physics associated with current and near-future facilities, such as DESI, Rubin LSST, and CMB-S4, is extremely rich. Dark matter science should be supported within these projects on intermediate scales in parallel to studies of dark energy and inflation. Such a program will fully leverage the unprecedented capabilities of these facilities. 
On large scales, the construction of future cosmology experiments is critical for expanding our understanding of dark matter physics. HEP involvement will be essential for the design and construction of these facilities, and dark matter physics should be a core component of their scientific mission. 

Cosmic probes of dark matter are unique and important, because they have sensitivity to microscopic physics of dark matter and provide precision measurements, regardless of whether dark matter has sizable interactions with normal matter. The HEP community has recognized the power of this approach, and it is now time to maximize its full potential. The support for comic probes, which may be the only viable way to measure dark matter properties, is essential for the decade of dark matter to come. 





\begingroup
\let\cleardoublepage\clearpage
\bibliographystyle{JHEP}
\bibliography{Cosmic/CF03/myreferences.bib}
\endgroup

\end{document}